\documentclass[10pt,5p,twocolumn]{elsarticle}

\usepackage[english]{babel}

\usepackage{graphicx} 
\usepackage{caption}
\usepackage[font=small]{caption}
\usepackage{subcaption}
\usepackage{booktabs}
\usepackage{amsmath}
\usepackage{slashed}

\usepackage{axodraw4j}
\usepackage{sectsty}
\allsectionsfont{\sffamily\mdseries\upshape}

\usepackage[unicode=true,ps2pdf]{hyperref}
\hypersetup{pdftitle=Rare decay \textbackslash pi\^{}0 -> e\^{}+e\^{}-: on 
corrections beyond the leading order}

\begin{document}

\begin{frontmatter}

\title{Rare decay $\pi^0\to e^+e^-$: on corrections beyond the leading order}
\author{Tom\'{a}\v{s} Husek}\ead{husek@ipnp.mff.cuni.cz}
\author{Karol Kampf}\ead{karol.kampf@mff.cuni.cz}
\author{Ji\v{r}\'{i} Novotn\'{y}}\ead{jiri.novotny@mff.cuni.cz}
\address{Institute of Particle and Nuclear Physics, Faculty of Mathematics and 
Physics, Charles University\\V Hole\v{s}ovi\v{c}k\'{a}ch 2, Praha 8, 
Czech Republic}

\begin{abstract}
The preceding experimental and theoretical results on the rare decay 
$\pi^0\to e^+e^-$ are briefly summarized. Already computed two-loop 
QED corrections are reviewed and the bremsstrahlung contribution beyond the 
soft-photon approximation is analytically calculated. The possible 
further contribution of QCD loop corrections is estimated using the leading 
logarithm approximation. 
The complete result can be used to fit the value of the contact interaction 
coupling $\chi^\text{(r)}$ to the recent KTeV experiment with the result 
$\chi^\text{(r)}(M_\rho)=4.5\pm1.0$.
\end{abstract}


\end{frontmatter}

%=======================================================================================================================================================================
%=======================================================================================================================================================================
%Motivation
%=======================================================================================================================================================================
%=======================================================================================================================================================================

\section{Motivation}

Experimental measurements of the rare decay of a neutral pseudoscalar meson to 
a lepton pair and its comparison with theoretical predictions offer an 
interesting way to study low-energy (long-distance) dynamics in the Standard 
Model (SM) \cite{Dorokhov:2007bd,Dorokhov:2009dg,Vasko:2011pi}. Systematical 
theoretical treatment of the process dates back to 1959, when the first 
prediction of the decay rate was published by Drell \cite{Drell}. While the 
possible contributions of the weak sector of the SM are small enough to be 
neglected, the leading order QED contribution is described by two virtual photon 
exchange triangle diagram. That is why the double off-shell pion transition form 
factor $F_{\pi^0\gamma^*\gamma^*}$, which is not known from the first 
principles, plays essential role.

Because of this one-loop structure for the leading order, the process is very 
rare and suppressed in the comparison to two photons decay 
($\pi^0\rightarrow\gamma\gamma$) by a factor of 2$\left({\alpha 
m_e}/{M_{\pi^0}}\right)^2$ due to the approximate helicity conservation of the 
interaction and thus may be sensitive to possible effects of the physics beyond 
the SM (expected branching ratio from the pure SM calculation is about 
$10^{-7}$).

Recently, this decay has attracted attention of the theorists again in 
connection with a new precise branching ratio measurement. The KTeV-E799-II 
experiment at Fermilab \cite{Abouzaid:2006kk} has observed $\pi^0\to e^+e^-$ 
events (altogether 794 candidates), where $K_L\rightarrow3\pi^0$ decay was used 
as a source of neutral pions. The KTeV result is
\begin{equation}
\begin{split}
\frac{\Gamma(\pi^0\to e^+e^-,\,x>0.95)}{\Gamma(\pi^0\rightarrow e^+e^-\gamma,\,x>0.232)}&=\\
=(1.685\pm0.064\pm&0.027)\times10^{-4}\,.
\end{split}
\label{eq:origexp}
\end{equation}
Here we have introduced the Dalitz variable
\begin{equation}
x
\equiv\frac{(p+q)^2}{M^2}
=\frac{(P-k)^2}{M^2}
=1-\frac{2E_k}{M}\,,
\label{eq:xEk}
\end{equation}
where $p$, $q$ and $k$ are four-momenta of electron, positron and photon, 
respectively, $P=(p+q+k)$ is the four-momentum of neutral pion $\pi^0$ with a 
mass $M$ and $E_k$ is the energy of the real outgoing photon in the pion CMS. 
The lower bound of the Dalitz variable $x$ is used to suppress the contribution 
of the Dalitz decay $\pi^0\to e^+e^-\gamma$, which naturally arises with lower 
$x$.

By means of extrapolating the Dalitz branching ratio in (\ref{eq:origexp}) to 
the full range of $x$, the branching ratio of the neutral pion decay into an 
electron-positron pair was determined to be equal to
\begin{equation}
\begin{split}
B(\pi^0\to e^+e^-(\gamma),\,x>0.95)&=\\
=(6.44\pm0.25\pm&0.22)\times10^{-8}\,.
\end{split}
\label{eq:B}
\end{equation}
Here the first error is from data statistics alone and the second is the total 
systematic error. For the matter of interest, current PDG average value 
$(6.46 \pm 0.33)\times 10^{-8}$~\cite{PDG} is mainly based on this new 
result. 

The KTeV collaboration used the result (\ref{eq:B}) for further calculations. 
They used the early calculation of Bergstr\"{o}m \cite{Bergstrom:1982wk} to 
extrapolate the full radiative tail beyond $x>0.95$ and to scale the result back 
up by the overall radiative corrections of 3.4\,\% to get the lowest order rate 
(with the final state radiation removed) for $\pi^0\to e^+e^-$ process. The 
final result is
\begin{equation}
B^{\text{no-rad}}_{\text{KTeV}}(\pi^0\to 
e^+e^-)=(7.48\pm0.29\pm0.25)\times10^{-8}\,.
\end{equation}

Subsequent comparison with theoretical predictions of the SM was made in 
\cite{Dorokhov:2007bd,Dorokhov:2009dg} using pion transition form factor data 
from CELLO \cite{Behrend:1990sr} and CLEO \cite{Gronberg:1997fj} experiments. 
Finally, it has been found, that according to SM the result should be
\begin{equation}
B^{\text{no-rad}}_{\text{SM}}(\pi^0\to e^+e^-)=(6.23\pm0.09)\times10^{-8}\,.
\end{equation}
This can be interpreted as a 3.3\,$\sigma$ discrepancy between the theory and the experiment. Of course, the discrepancy initiated further theoretical investigation of its possible sources~\cite{Dorokhov:2008qn,Kahn:2007ru}. Aside from the attempts to find the corresponding mechanism within the physics beyond the SM, also the possible revision of the SM predictions has been taken into account. Many corrections of this kind have been already made, but so far with no such a significant influence on the final result.

%=======================================================================================================================================================================
%=======================================================================================================================================================================
%Leading order
%=======================================================================================================================================================================
%=======================================================================================================================================================================

\section{Leading order}

\begin{figure}[t]
\centering
\setlength{\unitlength}{0.40pt}
\begin{picture}(630,144) (161,-148)
    \SetScale{0.40}
    \SetWidth{1.0}
    \SetColor{Black}
    \Line[dash,dashsize=10](636,-83)(699,-83)
    \Line[arrow,arrowpos=0.5,arrowlength=5,arrowwidth=2,arrowinset=0.2](699,-84)(771,-12)
    \Line[arrow,arrowpos=0.5,arrowlength=5,arrowwidth=2,arrowinset=0.2](771,-155)(699,-83)
    \Text(580,-100)[lb]{\huge{\Black{$+$}}}
    \Text(346,-95)[lb]{\LARGE{\Black{$=$}}}
    \Text(685,-116)[lb]{\normalsize{\Black{$\chi$}}}
    \Line[dash,dashsize=10](405,-83)(459,-83)
    \Photon(459,-83)(531,-29){4}{5.5}
    \Line[arrow,arrowpos=0.5,arrowlength=5,arrowwidth=2,arrowinset=0.2](561,-161)(531,-137)
    \Line[arrow,arrowpos=0.5,arrowlength=5,arrowwidth=2,arrowinset=0.2](531,-137)(531,-29)
    \Line[arrow,arrowpos=0.5,arrowlength=5,arrowwidth=2,arrowinset=0.2](531,-29)(561,-5)
    \Photon(531,-137)(459,-83){4}{5.5}
    \Line[dash,dashsize=10](172,-84)(226,-84)
    \Photon(226,-84)(298,-30){4}{5.5}
    \Line[arrow,arrowpos=0.5,arrowlength=5,arrowwidth=2,arrowinset=0.2](328,-162)(298,-138)
    \Line[arrow,arrowpos=0.5,arrowlength=5,arrowwidth=2,arrowinset=0.2](298,-138)(298,-30)
    \Line[arrow,arrowpos=0.5,arrowlength=5,arrowwidth=2,arrowinset=0.2](298,-30)(328,-6)
    \Photon(298,-138)(226,-84){4}{5.5}
    \GOval(244,-84)(18,18)(0){0.882}
    \CBox(694,-89)(706,-77){Black}{Black}
\end{picture}
\caption{Leading order contribution in the QED expansion and its representation in terms of the leading order of the chiral perturbation theory.}
\label{fig:LO}
\end{figure}
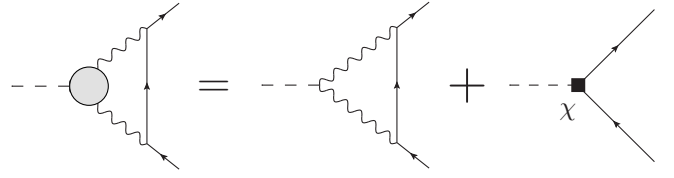

According to the Lorentz symmetry the on-shell invariant matrix element of the $\pi^0\to e^+e^-$ process can be generally written in terms of just one pseudoscalar form factor
\begin{equation}
i\mathcal{M}(\pi^0\to e^+e^-)
=\overline{u}(p,m)\gamma^{5}v(q,m)P(p^2,q^2,P^2)
\end{equation}
and, as a consequence, the total decay rate is given by
\begin{equation}
\Gamma(\pi^0\to e^+e^-)
=\frac M{8\pi}\sqrt{1-\nu^2}\left|P(m^2,m^2,M^2)\right|^2\,,
\label{rate}
\end{equation}
where $m$ stands for electron mass and $\nu \equiv 2m/M$. The leading order in 
the QED expansion is depicted as the left hand side of the graphical equation in 
the Fig.~\ref{fig:LO}. Here the shaded blob corresponds to the off-shell pion 
transition form factor $F_{\pi^0\gamma^*\gamma^*}(l^2,(P-l)^2)$ where $l$ is the 
loop momentum. This form factor serves as an effective UV cut-off due to its 
$1/l^2$ asymptotics governed by OPE (see e.g.~\cite{Knecht:2001xc}) and the loop 
integral over $\text{d}^4l$ is therefore convergent. It is convenient to pick up 
explicitly the non-analytic contribution of the two-photon intermediate state 
(the imaginary part\footnote{Imaginary part of this contribution is given by 
Cutkosky rules cutting the two virtual photon lines in the Fig.~\ref{fig:LO}.}
is determined uniquely up to the normalization given by the on-shell value of 
$F_{\pi^0\gamma^*\gamma^*}(0,0)\equiv F_{\pi^0\gamma\gamma}$) and express the 
form factor in the following way (cf. \cite{Knecht:1999gb})
\begin{equation}
\begin{split}
&P^\text{LO}(m^2,m^2,M^2)\\
&=\alpha^2mF_{\pi^0\gamma\gamma}\frac 1{\sqrt{1-\nu^2}}\left[\text{Li}_2(z)-\text{Li}_2\left(\frac 1z\right)+i\pi\log(-z)\right]\\
&+2\alpha^2mF_{\pi^0\gamma\gamma}\bigg\{\frac 32\log\left(\frac{m^2}{\mu^2}\right)-\frac 52+\chi\left(\frac{M^2}{\mu^2},\,\frac{m^2}{\mu^2}\right)\bigg\}\,.
\end{split}
\label{eq:P}
\end{equation}
Here, $\text{Li}_2$ is the dilogarithm,
\begin{equation}
z=-\frac{1-\sqrt{1-\nu^2}}{1+\sqrt{1-\nu^2}}
\end{equation}%
and $\mu$ represents the intrinsic scale connected with the form factor%
\footnote{It means the scale at which the loop integral is effectively cut off. 
The term $\frac 32\log \left( m^2/\mu^2\right)$ represents the leading 
dependence of the form factor $P$ on this scale.} %
$F_{\pi^0\gamma^*\gamma^*}$. The function $\chi\left(P^2/\mu^2,\,m^2/\mu^2\right)$ represents the remainder which collects the contributions of higher intermediate states and is real and analytic%
\footnote{Note that the higher intermediate states, which appear when also the blob in the Fig.~\ref{fig:LO} is cut, start for $P^2\sim\mu^2$.} %
for $P^2/\mu^2<1$.

The leading order terms in the chiral expansion of the form factor 
$P^\text{LO}$ are depicted as the right hand side of the graphical equation in 
Fig.~\ref{fig:LO}. The $\pi^0\gamma\gamma$ vertex in the loop graph is local and 
corresponds to the leading order term of the chiral expansion of the form factor 
$F_{\pi^0\gamma^*\gamma^*}$. Therefore the loop integration is no more UV finite 
and a counterterm (represented by the tree graph in the Fig.~\ref{fig:LO}) is 
necessary. The sum of these two terms can be written in the form (\ref{eq:P}), 
where the transition form factor $F_{\pi^0\gamma\gamma}$ and the remainder 
$\chi\left(P^2/\mu^2,m^2/\mu^2\right)$ are replaced by their leading orders in 
the chiral expansion
\begin{equation}
F_{\pi^0\gamma\gamma}^\text{LO}
=\frac 1{4\pi^2F}\,,\;\chi^\text{LO}\left(P^2/\mu^2,m^2/\mu^2\right)
=\chi^{\text{(r)}}(\mu)\,,
\label{eq:chiLO}
\end{equation}
where $\chi^\text{(r)}(\mu)$ is the finite part of the above mentioned 
counterterm renormalized at scale $\mu$. The graphical equation in the 
Fig.~\ref{fig:LO} can be understood as the matching condition for 
$\chi^\text{(r)}(\mu)$ at the leading order in the chiral expansion. It enables 
to determine $\chi^\text{(r)}(\mu)$ once the form factor 
$F_{\pi^0\gamma^*\gamma^*}$ is known. The latter can be theoretically modeled 
e.g. by the lowest meson dominance (LMD) approximation to the large-$N_{C}$ 
spectrum of vector meson resonances yielding~\cite{Knecht:1999gb}
\begin{equation}
\chi^\text{(r)}(M_\rho)=2.2\pm0.9\,,
\label{chi22}
\end{equation}
where $M_\rho=770\,\text{MeV}$ is the mass of the $\rho$ meson. For other 
alternative estimates cf. Tab.~\ref{tab:chi} and for the 
complete discussion see \cite{Dorokhov:2007bd}.

\begin{table}[!ht]
\begin{center}
{\scriptsize
\begin{tabular}{c | c c c c}
\toprule
Model & \text{CLEO+OPE} & \text{QCDsr} & \text{LMD+V} & \text{N$\chi $QM}\\
\midrule
$\chi^\text{(r)}(M_\rho)$ & $2.6\pm0.3$ & $2.8\pm0.1$ & 2.5 &
$2.4\pm0.5$\\
\bottomrule
\end{tabular}}
\end{center}
\caption{Numerical values of $\chi^\text{(r)}$ in different models 
according to \cite{Dorokhov:2007bd,Vasko:2011pi}. The first two columns 
denoted as CLEO+OPE and QCDsr correspond to various treatments of CLEO data. 
LMD+V is an improvement of the LMD ansatz and N$\chi$QM stands for the nonlocal 
chiral quark model.}
\label{tab:chi}
\end{table}

Using the value (\ref{chi22}) we get for the $\pi^0\to e^+e^-$ branching 
ratio numerically
\begin{equation}
B_\text{SM}^\text{LO}(\pi^0\to e^+e^-)
=(6.1\pm 0.3)\times10^{-8}\,.
\end{equation}

%=======================================================================================================================================================================
%=======================================================================================================================================================================
%Two-loop
%=======================================================================================================================================================================
%=======================================================================================================================================================================

\section{Two-loop virtual radiative corrections}

The full two-loop virtual radiative (pure QED) corrections of order 
$\mathcal{O}(\alpha^3p^2)$ were calculated 
in~\cite{Vasko:2011pi}. In this section we will present a short review of the 
main results. 

\begin{figure}[t]
\centering

\begin{subfigure}[t]{0.5\columnwidth}
\centering
\setlength{\unitlength}{0.45pt}
\begin{picture}(182,182) (179,-179)
    \SetScale{0.45}
    \SetWidth{1.0}
    \SetColor{Black}
    \Line[dash,dashsize=10](180,-88)(234,-88)
    \Photon(234,-88)(306,-34){4}{5.5}
    \Line[arrow,arrowpos=0.5,arrowlength=5,arrowwidth=2,arrowinset=0.2](360,-178)(306,-142)
    \Line[arrow,arrowpos=0.5,arrowlength=5,arrowwidth=2,arrowinset=0.2](306,-142)(306,-34)
    \Photon(306,-142)(234,-88){4}{5.5}
    \Line[arrow,arrowpos=0.3,arrowlength=5,arrowwidth=2,arrowinset=0.2](306,-34)(360,2)
    \PhotonArc(319.5,-38.5)(26.239,-120.964,59.036){4}{5.5}
\end{picture}
\caption{}
\label{fig:2a}
\end{subfigure}
\begin{subfigure}[t]{0.4\columnwidth}
\centering
\setlength{\unitlength}{0.45pt}
\begin{picture}(182,182) (179,-179)
    \SetScale{0.45}
    \SetWidth{1.0}
    \SetColor{Black}
    \Line[dash,dashsize=10](180,-88)(234,-88)
    \Photon(234,-88)(306,-34){4}{5.5}
    \Line[arrow,arrowpos=0.7,arrowlength=5,arrowwidth=2,arrowinset=0.2](360,-178)(306,-142)
    \Line[arrow,arrowpos=0.5,arrowlength=5,arrowwidth=2,arrowinset=0.2](306,-142)(306,-34)
    \Photon(306,-142)(234,-88){4}{5.5}
    \Line[arrow,arrowpos=0.5,arrowlength=5,arrowwidth=2,arrowinset=0.2](306,-34)(360,2)
    \PhotonArc[clock](319.5,-137.5)(26.239,120.964,-59.036){4}{5.5}
\end{picture}
\caption{}
\label{fig:2b}
\end{subfigure}

\begin{subfigure}[t]{0.5\columnwidth}
\centering
\setlength{\unitlength}{0.45pt}
\begin{picture}(182,182) (179,-179)
    \SetScale{0.45}
    \SetWidth{1.0}
    \SetColor{Black}
    \Line[dash,dashsize=10](180,-88)(234,-88)
    \Photon(234,-88)(306,-34){4}{5.5}
    \Line[arrow,arrowpos=0.5,arrowlength=5,arrowwidth=2,arrowinset=0.2](360,-178)(306,-142)
    \Line[arrow,arrowpos=0.5,arrowlength=5,arrowwidth=2,arrowinset=0.2](306,-142)(306,-34)
    \Photon(306,-142)(234,-88){4}{5.5}
    \Line[arrow,arrowpos=0.5,arrowlength=5,arrowwidth=2,arrowinset=0.2](306,-34)(360,2)
    \PhotonArc[clock](313.875,-88)(28.125,106.26,-106.26){4}{6.5}
\end{picture}
\caption{}
\label{fig:2c}
\end{subfigure}
\begin{subfigure}[t]{0.4\columnwidth}
\centering
\setlength{\unitlength}{0.45pt}
\begin{picture}(186,182) (179,-179)
    \SetScale{0.45}
    \SetWidth{1.0}
    \SetColor{Black}
    \Line[dash,dashsize=10](180,-88)(234,-88)
    \Photon(234,-88)(306,-34){4}{5.5}
    \Line[arrow,arrowpos=0.7,arrowlength=5,arrowwidth=2,arrowinset=0.2](360,-178)(306,-142)
    \Line[arrow,arrowpos=0.5,arrowlength=5,arrowwidth=2,arrowinset=0.2](306,-142)(306,-34)
    \Photon(306,-142)(234,-88){4}{5.5}
    \Line[arrow,arrowpos=0.3,arrowlength=5,arrowwidth=2,arrowinset=0.2](306,-34)(360,2)
    \PhotonArc[clock](250.5,-88)(109.5,41.112,-41.112){4}{8.5}
\end{picture}
\caption{}
\label{fig:2d}
\end{subfigure}

\begin{subfigure}[t]{0.5\columnwidth}
\centering
\setlength{\unitlength}{0.45pt}
\begin{picture}(182,182) (179,-179)
    \SetScale{0.45}
    \SetWidth{1.0}
    \SetColor{Black}
    \Line[dash,dashsize=10](180,-88)(234,-88)
    \Photon(288,-52)(306,-34){4}{1.5}
    \Line[arrow,arrowpos=0.5,arrowlength=5,arrowwidth=2,arrowinset=0.2](360,-178)(306,-142)
    \Line[arrow,arrowpos=0.5,arrowlength=5,arrowwidth=2,arrowinset=0.2](306,-142)(306,-34)
    \Line[arrow,arrowpos=0.5,arrowlength=5,arrowwidth=2,arrowinset=0.2](306,-34)(360,2)
    \Arc[arrow,arrowpos=0.5,arrowlength=5,arrowwidth=2,arrowinset=0.2](270,-61)(20.125,153,513)
    \Photon(234,-88)(253,-71){4}{1.5}
    \Photon(306,-142)(234,-88){4}{5.5}
\end{picture}
\caption{}
\label{fig:2e}
\end{subfigure}
\begin{subfigure}[t]{0.4\columnwidth}
\centering
\setlength{\unitlength}{0.45pt}
\begin{picture}(182,182) (179,-179)
    \SetScale{0.45}
    \SetWidth{1.0}
    \SetColor{Black}
    \Line[dash,dashsize=10](180,-88)(234,-88)
    \Line[arrow,arrowpos=0.5,arrowlength=5,arrowwidth=2,arrowinset=0.2](360,-178)(306,-142)
    \Line[arrow,arrowpos=0.5,arrowlength=5,arrowwidth=2,arrowinset=0.2](306,-142)(306,-34)
    \Photon(306,-142)(288,-124){4}{1.5}
    \Line[arrow,arrowpos=0.5,arrowlength=5,arrowwidth=2,arrowinset=0.2](306,-34)(360,2)
    \Arc[arrow,arrowpos=0.7,arrowlength=5,arrowwidth=2,arrowinset=0.2](270,-115)(20.125,153,513)
    \Photon(253,-104)(234,-88){4}{1.5}
    \Photon(234,-88)(306,-34){4}{5.5}
\end{picture}
\caption{}
\label{fig:2f}
\end{subfigure}
\caption{Two-loop virtual radiative corrections for $\pi^0\to e^+e^-$ process.}
\label{fig:2loop}
\end{figure}
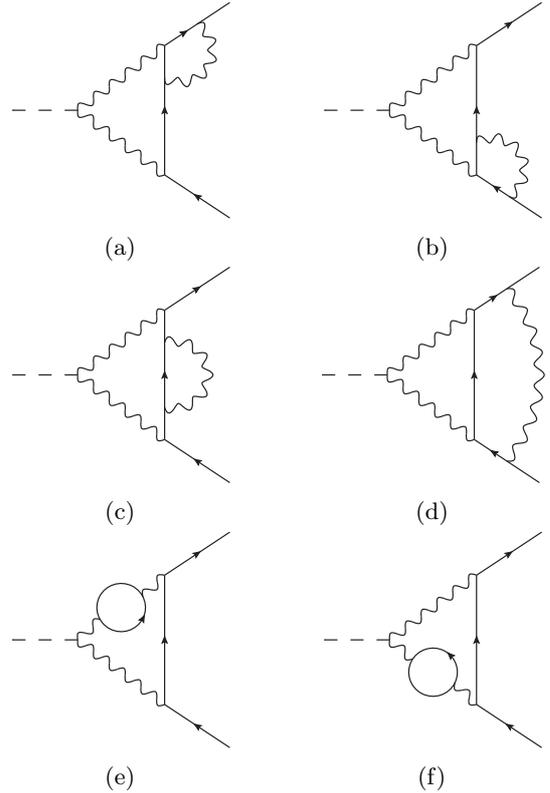

The relevant contributions to the amplitude are shown in Fig.~\ref{fig:2loop}. 
There are six two-loop diagrams. Listed sequentially, we have two vertex 
corrections (a, b), electron self-energy insertion (c), box-type correction (d) 
and two vacuum polarization insertions (e, f). Of course, for every such diagram 
a one-loop graph with corresponding counterterm must be added to renormalize 
the subdivergences. The relevant finite parts of these counterterms can be fixed 
by the requirement that the parameters $m$ and $\alpha$ coincide with their 
physical values. After the subdivergences are canceled, the remaining 
superficial divergences has to be renormalized by another additional tree 
counter-term with coupling $\xi$. %
The finite part $\xi^\text{(r)}(\mu)$ of this coupling has been estimated 
in~\cite{Vasko:2011pi} using its running with the renormalization scale as
\begin{equation}
\xi^\text{(r)}(M_\rho)=0\pm5.5\,.
\label{xir}
\end{equation}
Besides the UV divergences, the graph (d) in the Fig.~\ref{fig:2loop} is also 
IR divergent. It is therefore necessary to consider IR-safe decay width of the 
inclusive process $\pi^0\to e^+e^-(\gamma)$ with additional real photon in the 
final state. In~\cite{Vasko:2011pi} the real photon bremsstrahlung has been 
taken into account using the soft-photon approximation. The final result depends 
on the experimental upper bound on the soft photon energy which can be expressed 
in terms of the lower bound $x^\text{cut}$ on the Dalitz variable $x$ (see 
(\ref{eq:xEk})). The result can be expressed in terms of the correction factor 
$\delta(x^\text{cut})$ defined as
\begin{equation}
\begin{split}
&\Gamma^\text{NLO}\left(\pi^0\to e^+e^-(\gamma ),\,x>x^\text{cut}\right)\\
&\equiv\delta \left(x^\text{cut}\right)\Gamma^\text{LO}\left(\pi^0\to e^+e^-\right)\,,
\end{split}
\end{equation}
where $\Gamma^\text{LO}$ is the leading order width and $\Gamma^\text{NLO}$ is 
the next-to leading $\mathcal{O}(\alpha^3p^2)$ correction. The $x^\text{cut}$ 
dependent overall correction $\delta(x^\text{cut})$ has various sources and to 
emphasize the origin of its constituents, we will use the same symbol decorated 
with appropriate
indices. For the complete QED two-loop correction $\delta^{(2)}$ including soft-photon bremsstrahlung and KTeV cut $x^\text{cut}$=$0.95$, in~\cite{Vasko:2011pi} it was obtained
\begin{equation}
\delta^{(2)}(0.95)
\equiv\delta^{\text{virt.}}+\delta_\text{soft}^\text{BS}(0.95)=(-5.8\pm0.2)\,\%\,,
\label{eq:2looprad}
\end{equation}
where only the uncertainties of $\chi^\text{(r)}$ and $\xi^\text{(r)}$ were 
taken as the source of the error. This result differs significantly from the 
previous approximate calculations done by Bergstr\"{o}m~\cite{Bergstrom:1982wk} 
or Dorokhov et~al.~\cite{Dorokhov:2008qn}, where for 
$\delta^{(2)}\left(0.95\right) $ we would get $-13.8\,\%$ and $-13.3\,\%$, 
respectively.

There is a simple interrelation of this partial result of the QED radiative 
corrections and the branching ratio (\ref{eq:B}) obtained by KTeV experiment 
(for the details see~\cite{Vasko:2011pi}). We can write the theoretical 
prediction for the branching ratio measured by KTeV as
\begin{equation}
\begin{split}
&B(\pi^0\to e^+e^-(\gamma ),\,x>0.95)
=\frac{\Gamma^\text{LO}(\pi^0\to e^+e^-)}{\Gamma (\pi^0\to\gamma\gamma)}\\
\times\,&B(\pi^0\to \gamma\gamma)\left[1+\delta^{(2)}(0.95)+\Delta^\text{BS}(0.95)+\delta^\text{D}(0.95)\right]\,,
\end{split}
\label{eq:Bth}
\end{equation}
where the only experimental input is the precise branching ratio $B(\pi^0\to\gamma\gamma)=(98.823\pm0.034)$\,\%. In the above formula,
\begin{align}
\delta^{\text{D}}(x^\text{cut})
&=\frac1{\Gamma^{\text{LO}}(\pi^0\to e^+e^-)}
{\int_{x^\text{cut}}^1}{{\text{d}}x}\left(\frac{\text{d}\Gamma^{\text{Dalitz}}}{\text{d}x}\right)_{1\gamma IR}^{\text{NLO}}\notag\\
&=\frac{1.75\times 10^{-15}}{\left[\Gamma^\text{LO}(\pi^0\to e^+e^-)/\text{MeV}\right]}
\end{align}
corresponds to the unsubtracted fraction of the Dalitz decay background%
\footnote{This fraction comes form the contribution of the interference term of 
the NLO one-photon-irreducible ($1\gamma IR$) graph with the leading order 
Dalitz amplitude. See \cite{Vasko:2011pi} and \cite{Kampf:2005tz} for more 
details.} %
omitted in the KTeV analysis and discussed in~\cite{Kampf:2005tz,Vasko:2011pi}. 
In what follows we will concentrate on the last missing ingredient of the 
formula (\ref{eq:Bth}), namely
\begin{equation}
\Delta^\text{BS}\left(x^\text{cut}\right)
\equiv\delta^\text{BS}\left(x^\text{cut}\right)-\delta_\text{soft}^\text{BS}\left(x^\text{cut}\right)\,,
\end{equation}
which is the difference between the exact bremsstrahlung and its soft photon approximation. This difference has been only roughly estimated in~\cite{Vasko:2011pi} and this estimate has been taken as a source of the error. Our aim is to calculate $\Delta^{\text{BS}}$ exactly and test the adequacy of the soft photon approximation for the cut $x^\text{cut}=0.95$ used in the KTeV analysis.

%=======================================================================================================================================================================
%=======================================================================================================================================================================
%Bremsstrahlung
%=======================================================================================================================================================================
%=======================================================================================================================================================================

\section{Bremsstrahlung}
\label{sec:bs}

\begin{figure}[t]
\centering
\begin{subfigure}[b]{0.4\columnwidth}
\centering
\setlength{\unitlength}{0.5pt}
\begin{picture}(202,217) (190,-144)
    \SetScale{0.5}
    \SetWidth{1.0}
    \SetColor{Black}
%    \Line[arrow,arrowpos=1,arrowlength=5,arrowwidth=2,arrowinset=0.2](243,-35)(279,-8)
    \Line[dash,dashsize=10](170,-53)(234,-53)
%    \Text(170,-43)[lb]{\tiny{\Black{$\pi^0\left(P\right)$}}}
%    \Text(350,52)[lb]{\tiny{\Black{$e^-\left(p\right)$}}}
%    \Text(340,-123)[lb]{\tiny{\Black{$e^+\left(-q\right)$}}}
    \Photon(234,-53)(306,1){4}{5.5}
    \Line[arrow,arrowpos=0.5,arrowlength=5,arrowwidth=2,arrowinset=0.2](360,-143)(306,-107)
    \Line[arrow,arrowpos=0.5,arrowlength=5,arrowwidth=2,arrowinset=0.2](306,-107)(306,1)
    \Line[arrow,arrowpos=0.5,arrowlength=5,arrowwidth=2,arrowinset=0.2](342,25)(378,49)
    \Photon(306,-107)(234,-53){4}{5.5}
%    \Text(320,-58)[lb]{\tiny{\Black{$p+k-l$}}}
%    \Text(255,-18)[lb]{\tiny{\Black{$\rput[lb]{37}{l}$}}}
%    \Text(225,-77)[lb]{\tiny{\Black{$\rput[lb]{-37}{p+q+k-l}$}}}
%    \Line[arrow,arrowpos=1,arrowlength=5,arrowwidth=2,arrowinset=0.2](243,-71)(279,-98)
%    \Line[arrow,arrowpos=1,arrowlength=5,arrowwidth=2,arrowinset=0.2](315,-73)(315,-33)
    \Photon(342,25)(378,1){4}{3.5}
    \Line[arrow,arrowpos=0.5,arrowlength=5,arrowwidth=2,arrowinset=0.2](306,1)(342,25)
%    \Text(306,19)[lb]{\tiny{\Black{$\rput[lb]{34}{p+k}$}}}
%    \Line[arrow,arrowpos=1,arrowlength=5,arrowwidth=2,arrowinset=0.2](308,13)(332,29)
%    \Text(350,-18)[lb]{\tiny{\Black{$\gamma\left(k\right)$}}}
\end{picture}
\caption{}
\label{fig:T2}
\end{subfigure}
\begin{subfigure}[b]{0.45\columnwidth}
\centering
\setlength{\unitlength}{0.5pt}
\begin{picture}(225,206) (155,-179)
    \SetScale{0.5}
    \SetWidth{1.0}
    \SetColor{Black}
%    \Line[arrow,arrowpos=1,arrowlength=5,arrowwidth=2,arrowinset=0.2](243,-46)(279,-19)
    \Line[dash,dashsize=10](170,-64)(234,-64)
%    \Text(170,-54)[lb]{\tiny{\Black{$\pi^0\left(P\right)$}}}
%    \Text(340,-14)[lb]{\tiny{\Black{$e^-\left(p\right)$}}}
%    \Text(348,-184)[lb]{\tiny{\Black{$e^+\left(-q\right)$}}}
    \Photon(234,-64)(306,-10){4}{5.5}
    \Line[arrow,arrowpos=0.5,arrowlength=5,arrowwidth=2,arrowinset=0.2,flip](360,26)(306,-10)
    \Line[arrow,arrowpos=0.5,arrowlength=5,arrowwidth=2,arrowinset=0.2](306,-118)(306,-10)
    \Line[arrow,arrowpos=0.5,arrowlength=5,arrowwidth=2,arrowinset=0.2,flip](342,-142)(378,-166)
    \Photon(306,-118)(234,-64){4}{5.5}
%    \Text(320,-69)[lb]{\tiny{\Black{$q+k-l$}}}
%    \Text(235,-49)[lb]{\tiny{\Black{$\rput[lb]{37}{p+q+k-l}$}}}
%    \Text(252,-106)[lb]{\tiny{\Black{$\rput[lb]{-37}{l}$}}}
%    \Line[arrow,arrowpos=1,arrowlength=5,arrowwidth=2,arrowinset=0.2](243,-82)(279,-109)
%    \Line[arrow,arrowpos=0,arrowlength=5,arrowwidth=2,arrowinset=0.2,flip](315,-84)(315,-44)
    \Photon(342,-142)(378,-118){4}{3.5}
    \Line[arrow,arrowpos=0.5,arrowlength=5,arrowwidth=2,arrowinset=0.2,flip](306,-118)(342,-142)
%    \Text(303,-145)[lb]{\tiny{\Black{$\rput[lb]{-34}{q+k}$}}}
%    \Line[arrow,arrowpos=1,arrowlength=5,arrowwidth=2,arrowinset=0.2](308,-130)(332,-146)
%    \Text(348,-112)[lb]{\tiny{\Black{$\gamma\left(k\right)$}}}
\end{picture}
\caption{}
\label{fig:T3}
\end{subfigure}
\begin{subfigure}[b]{0.4\columnwidth}
\centering
\setlength{\unitlength}{0.5pt}
\begin{picture}(129,146) (196,-170)
    \SetScale{0.5}
    \SetWidth{1.0}
    \SetColor{Black}
    \Line[dash,dashsize=10](180,-97)(251,-97)
    \Line[arrow,arrowpos=0.5,arrowlength=5,arrowwidth=2,arrowinset=0.2](251,-98)(297,-52)
    \Line[arrow,arrowpos=0.5,arrowlength=5,arrowwidth=2,arrowinset=0.2](323,-169)(251,-97)
    \Text(237,-130)[lb]{\normalsize{\Black{$\chi$}}}
    \Line[arrow,arrowpos=0.5,arrowlength=5,arrowwidth=2,arrowinset=0.2](297,-52)(324,-25)
    \Photon(297,-52)(324,-70){4}{2.5}
    \CBox(246,-104)(258,-92){Black}{Black}
\end{picture}
\caption{}
\label{fig:T2CT}
\end{subfigure}
\begin{subfigure}[b]{0.45\columnwidth}
\centering
\setlength{\unitlength}{0.5pt}
\begin{picture}(129,146) (196,-170)
    \SetScale{0.5}
    \SetWidth{1.0}
    \SetColor{Black}
    \Line[dash,dashsize=10](180,-97)(251,-97)
    \Line[arrow,arrowpos=0.5,arrowlength=5,arrowwidth=2,arrowinset=0.2](251,-98)(297,-142)
    \Line[arrow,arrowpos=0.5,arrowlength=5,arrowwidth=2,arrowinset=0.2](324,-25)(251,-97)
    \Text(237,-130)[lb]{\normalsize{\Black{$\chi$}}}
    \Line[arrow,arrowpos=0.5,arrowlength=5,arrowwidth=2,arrowinset=0.2](297,-142)(324,-169)
    \Photon(297,-142)(324,-124){4}{2.5}
    \CBox(246,-104)(258,-92){Black}{Black}
\end{picture}
\caption{}
\label{fig:T3CT}
\end{subfigure}
\begin{subfigure}[b]{\columnwidth}
\centering
\setlength{\unitlength}{0.5pt}
\begin{picture}(257,182) (165,-179)
    \SetScale{0.5}
    \SetWidth{1.0}
    \SetColor{Black}
%    \Line[arrow,arrowpos=1,arrowlength=5,arrowwidth=2,arrowinset=0.2](243,-70)(279,-43)
    \Line[dash,dashsize=10](170,-88)(234,-88)
%    \Text(170,-78)[lb]{\scriptsize{\Black{$\pi^0\left(P\right)$}}}
%    \Text(340,-33)[lb]{\scriptsize{\Black{$e^-\left(p\right)$}}}
%    \Text(340,-158)[lb]{\scriptsize{\Black{$e^+\left(-q\right)$}}}
    \Photon(234,-88)(306,-34){4}{5.5}
    \Line[arrow,arrowpos=0.5,arrowlength=5,arrowwidth=2,arrowinset=0.2](360,-178)(306,-142)
    \Line[arrow,arrowpos=0.5,arrowlength=5,arrowwidth=2,arrowinset=0.2](306,-88)(306,-34)
    \Line[arrow,arrowpos=0.5,arrowlength=5,arrowwidth=2,arrowinset=0.2](306,-34)(360,2)
    \Photon(306,-142)(234,-88){4}{5.5}
%    \Text(320,-63)[lb]{\scriptsize{\Black{$p-l$}}}
%    \Text(255,-53)[lb]{\scriptsize{\Black{$\rput[lb]{37}{l}$}}}
%    \Text(230,-113)[lb]{\scriptsize{\Black{$\rput[lb]{-37}{k+p+q-l}$}}}
%    \Line[arrow,arrowpos=1,arrowlength=5,arrowwidth=2,arrowinset=0.2](243,-106)(279,-133)
%    \Line[arrow,arrowpos=1,arrowlength=5,arrowwidth=2,arrowinset=0.2](315,-78)(315,-43)
    \Line[arrow,arrowpos=0.5,arrowlength=5,arrowwidth=2,arrowinset=0.2](306,-142)(306,-88)
    \Photon(306,-88)(366,-88){4}{4}
%    \Line[arrow,arrowpos=1,arrowlength=5,arrowwidth=2,arrowinset=0.2](315,-138)(315,-103)
%    \Text(320,-125)[lb]{\scriptsize{\Black{$k+p-l$}}}
%    \Text(365,-83)[lb]{\scriptsize{\Black{$\gamma\left(k\right)$}}}
\end{picture}
\caption{}
\label{fig:T4}
\end{subfigure}
\caption{Bremsstrahlung Feynman diagrams for $\protect\pi^0\to e^+e^-$ process including counterterms.}
\label{fig:BSout}
\end{figure}
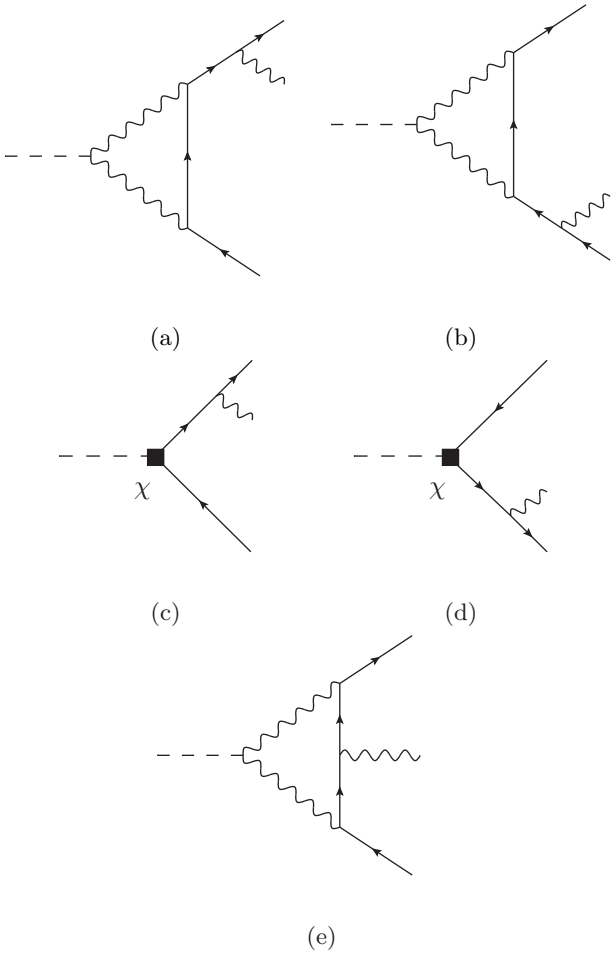

In this section, we discuss the above mentioned exact bremsstrahlung (BS), i.e. 
the real radiative correction corresponding to the process $\pi^0\to 
e^+e^-(\gamma)$ beyond the soft-photon approximation. As a consequence of the 
gauge invariance, the invariant amplitude for the BS correction
\begin{equation}
\mathcal{M}_{(\lambda )}(p,q,k)
\equiv \varepsilon _{(\lambda )}^{* \rho }(k)\mathcal{M}_{\rho }^{\text{BS}}(p,q,k)
\end{equation}
(where $k$ and $\varepsilon _{(\lambda )}^{* \rho}(k)$ is the photon momentum 
and polarization vector, respectively) has to satisfy the Ward identity
\begin{equation}
k^{\rho }\mathcal{M}_{\rho }^{\text{BS}}=0
\label{eq:BSWard}
\end{equation}
for on-shell $k$ and thus it can be generally expressed in the form~\cite{Kampf:2005tz}
\begin{equation}
\begin{split}
i&\mathcal{M}_{\rho }^{\text{BS}}(p,q,k)=\frac{i e^5}{8\pi^2 F}\times\\
&\;\;\Big\{P(x,y)\left[\left(k\cdot p\right) q_\rho-\left(k\cdot q\right) p_\rho\right]\left[\bar{u}(p,m)\gamma_5 v(q,m)\right]\\
&+A(x, y)\Big[\bar{u}(p,m)\left[\gamma_\rho\left(k\cdot p\right)-p_\rho (k\cdot\gamma)\right]\gamma_5 v(q,m)\Big]\\
&-A(x,-y)\Big[\bar{u}(p,m)\left[\gamma_\rho\left(k\cdot q\right)-q_\rho (k\cdot\gamma)\right]\gamma_5 v(q,m)\Big]\\
&+T(x,y)\left[\bar{u}(p,m)\gamma_\rho\slashed k\gamma_5 v(q,m)\right]\Big\}
\end{split}
\label{eq:BStotampl}
\end{equation}
in terms of scalar form factors $P$, $A$ and $T$. These are functions of two independent kinematic variables $(x,y)$, defined~as
\begin{equation}
\begin{gathered}
x=\frac{(p+q)^2}{M^2}\,,\;\;y=-\frac{2}{M^2}\left[\frac{k\cdot(p-q)}{1-x}\right]\\
x\in[\nu^2,1]\,,\;\;y\in\left[-\sqrt{1-\frac{\nu^2}{x}},\sqrt{1-\frac{\nu^2}{x}}\;\right]\,.
\end{gathered}
\label{eq:xybound}
\end{equation}
As mentioned above, $x$ is the Dalitz variable (i.e. a normalized square of the 
total energy of $e^+e^-$ pair in their CMS) and $y$ has the meaning of a 
rescaled cosine of the angle included by the directions of outgoing photon and 
positron in the $e^+e^-$ CMS. The modulus squared of the amplitude has the 
form~\cite{Kampf:2005tz}
\begin{equation}
\begin{split}
&\overline{\left|\mathcal{M}^\text{BS}(x,y)\right|^2}
\equiv\sum_{\text{polarizations}}\left|\mathcal{M}_{(\lambda)}(p,q,
k)\right|^2=\\
&=\frac{16\pi\alpha^5}{F^2}
%=\frac{16\pi\alpha^5}{F^2}
\frac{M^4(1-x)^2}{8}\bigg\{
M^2\left[x(1-y^2)-\nu^2\right]\Big[xM^2\left|P\right|^2\\
&+2\nu M\operatorname{Re}\big\{P^*\left[A(x,y)+A(x,-y)\right]\big\}-4\operatorname{Re}\big\{P^*T\big\}\Big]\\
&+2M^2(x-\nu^2)(1-y)^2\left|A(x,y)\right|^2+(y\to -y)\\
%&\quad+\left.(1+y)^2\left|A(x,-y)\right|^2\right]\\
&-8\nu My(1-y)\operatorname{Re}\big\{A(x,y)T^*\big\}+(y\to -y)\\
%&\quad-(1+y)\left.\operatorname{Re}\big\{A(x,-y)T^*\big\}\right]\\
&-4\nu^2M^2y^2\operatorname{Re}\big\{A(x,y)A(x,-y)^*\big\}
+8(1-y^2)\left|T\right|^2\bigg\}
\end{split}
\label{eqMbar2}
\end{equation}
and using the variables $x$, $y$ the differential decay rate is
\begin{equation}
\text{d}\Gamma^\text{BS}(x,y)
=\frac{M}{(8\pi)^3}\overline{\left|\mathcal{M}^\text{BS}(x,y)\right|^2}(1-x)\,\text{d}x\,\text{d}y\,.
\label{eq:dGammaBS}
\end{equation}
To the amplitude $\mathcal{M}_{(\lambda)}(p,q,k)$ five Feynman diagrams 
contribute (cf. Fig.~\ref{fig:BSout}). Four of them correspond to the photon 
emission from the outgoing fermion lines (see 
Fig.~\ref{fig:T2}---\ref{fig:T3CT}). Naively, one would expect that only these 
four diagrams are necessary to consider since only they include IR divergences 
which are needed to cancel the IR divergences stemming from the virtual 
corrections 
(see graph\ (d) in the Fig.~\ref{fig:2loop} and the corresponding one-loop 
diagram 
with counterterm). However, this result would not be complete.
The reason is  that the Ward identity (\ref{eq:BSWard}) would be violated%
\footnote{Note that in the framework of the soft-photon approximation the sum 
of these four graphs satisfies the Ward identity by itself.}. %
Thus it is necessary to add the third (box) diagram (Fig.~\ref{fig:T4}, photon 
emitted from the inner fermion line) to fulfill this relation.

In the graphs (\ref{fig:T2}) and (\ref{fig:T3}) the $\pi\gamma\gamma$ vertex 
stems from the Wess-Zumino-Witten action \cite{Wess:1971yu,Witten:1983tw} and 
the remaining vertices correspond to standard QED Feynman rules. These graphs 
are UV divergent by power counting and have to be regularized. In what follows, 
we use the dimensional regularization. In order to bypass the problems with 
intrinsically four-dimensional objects like $\gamma_5$ and the Levi-Civita 
pseudo-tensor $\varepsilon^{\mu\nu\alpha\beta}$, we use its variant known as 
Dimensional Reduction%
\footnote{Note however, that in general case the regularization by dimensional 
reduction might spoil gauge invariance. In the case of our amplitude, we have
checked that the gauge invariance is preserved and the regularized amplitude 
has the general form (\ref{eq:BStotampl}).} %
(cf.~\cite{Frampton:1978ix}), which keeps the algebra of $\gamma$-matrices 
four-dimensional while the loop tensor integrals are regularized dimensionally 
and expressed in terms of the scalar one-loop integrals using the 
Passarino-Veltman reduction \cite{Passarino:1978jh}. Within this framework we 
first get rid of the Levi-Civita tensor using the four-dimensional identities, 
e.g.
\begin{equation}
\begin{split}
\varepsilon^{\alpha\beta\mu\nu}\gamma_\mu\gamma_\nu&=i\gamma_5\left[\gamma^\alpha,\gamma^\beta\right]\\
\varepsilon^{\alpha\beta\mu\nu}
\gamma_\mu\gamma_\rho\gamma_\nu&=2i\gamma_5\left(g^\alpha_\rho\gamma^\beta
-g^\beta_\rho\gamma^\alpha\right)\,,
\end{split}
\end{equation}
and then contract the reduced tensor integrals with the $\gamma$-matrix structures%
\footnote{According to the prescription \cite{Frampton:1978ix}, we take the metric tensors stemming from the Passarino-Veltman reduction effectively as four-dimensional.}. %
The contributions of the box diagram Fig.~\ref{fig:T4} turns out to be finite 
while the triangle diagrams Fig.~\ref{fig:T2} and Fig.~\ref{fig:T3} contain 
subdivergences which have to be renormalized by means of the tree graphs with 
counterterm corresponding to the coupling $\chi$ (see Fig.~\ref{fig:T2CT} 
and~\ref{fig:T3CT}).\ Summing all the relevant contributions and using the 
four-dimensional Dirac algebra, we get finally the form factors $P$, $A$, and 
$T$, the explicit form of which is summarized in~\ref{app:BS}.

\begin{figure}[t]
\includegraphics[width=\columnwidth]{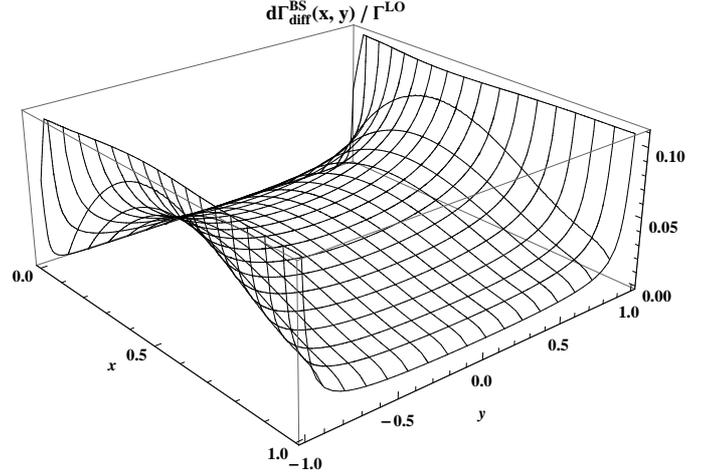}
\caption{3D plot of $\text{d}\Gamma_\text{diff}^\text{BS}(x,y)$ normalized to the leading order contribution of the $\protect\pi^0\to e^+e^-$ process.}
\label{fig:final}
\end{figure}

\begin{figure}[t]
\includegraphics[width=\columnwidth]{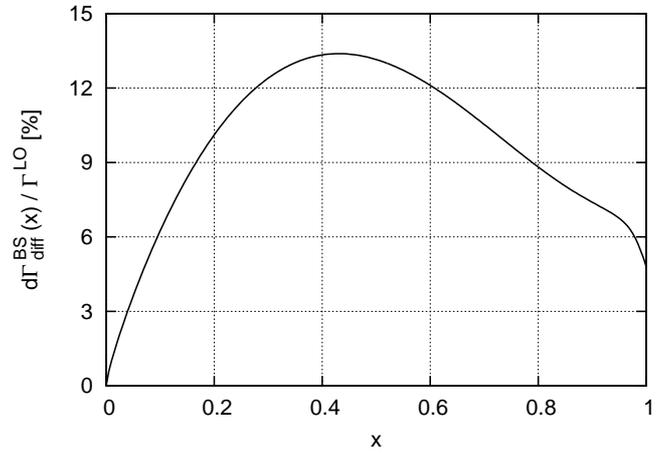}
\caption{Plot of $\protect\text{d}\Gamma_\text{diff}^\text{BS}(x)=\int\text{d}\Gamma_\text{diff}^\text{BS}(x,y)\,\text{d}y$ normalized to the leading order contribution of the $\protect\pi^0\to e^+e^-$ process.}
\label{fig:x2D}
\end{figure}

\begin{figure}[t]
\includegraphics[width=\columnwidth]{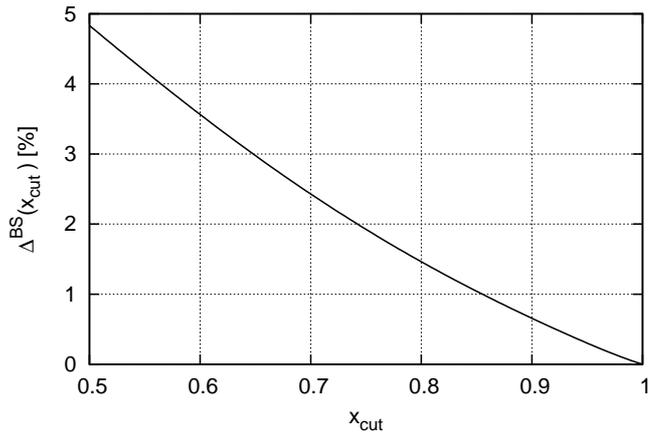}
\caption{The dependence of $\Delta^\text{BS}$ on the cut on the Dalitz variable.}
\label{fig:xcut2D}
\end{figure}

\begin{figure}[t]
\includegraphics[width=\columnwidth]{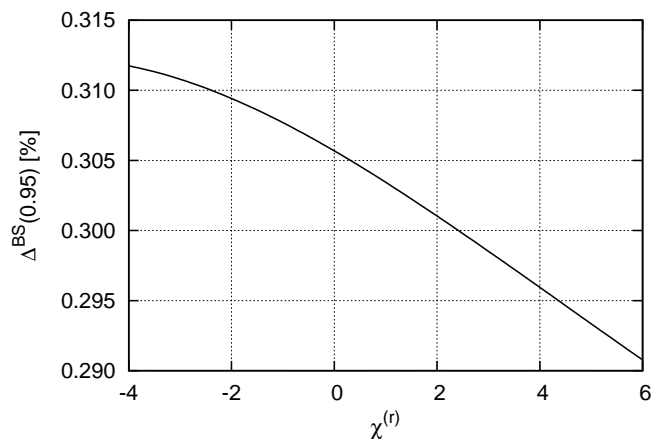}
\caption{The dependence of $\Delta^\text{BS}(0.95)$ on $\chi^\text{(r)}$. It is apparent, that the dependence is very slight and can be neglected in the calculation of the $\chi^\text{(r)}$.}
\label{fig:chi2D}
\end{figure}

The differential decay rate $\text{d}\Gamma^\text{BS}(x,y)$ (cf.~(\ref{eq:dGammaBS})) give rise to IR divergences when integrated over the phase space. The divergences originate from the soft-photon region
\begin{equation}
|\mathbf{k}|<\frac 12 M(1-x^\text{cut})\,,
\end{equation}%
which is defined in terms of the variables $(x,y)$ by means of the cut on the Dalitz variable $x>x^{\text{cut}}$. These divergences are exactly the same as those stemming from an analogous integral of the differential decay rate ${\text{d}}\Gamma _{\text{soft}}^{\text{BS}}(x,y)$ calculated within the soft-photon approximation. The latter is already included in the two-loop result \cite{Vasko:2011pi}, we therefore present our result for the exact BS as a difference
\begin{equation}
\text{d}\Gamma_\text{diff}^\text{BS}(x,y)
=\text{d}\Gamma^\text{BS}(x,y)-\text{d}\Gamma_\text{soft}^\text{BS}(x,y)\,,
\end{equation}
the integral of which is IR finite. The result for $\text{d}\Gamma_\text{diff}^\text{BS}(x,y)$ is shown in Fig.~\ref{fig:final} and (integrated over the allowed region of $y$ given by (\ref{eq:xybound})) in Fig.~\ref{fig:x2D}. For $\Delta ^{\text{BS}}(x^{\text{cut}})$ we get finally
\begin{equation}
\Delta^\text{BS}(x^\text{cut})
=2\int_{x^\text{cut}}^1\int_0^{\sqrt{1-\nu^2/x}}\frac{\text{d}\Gamma_\text{diff}^\text{BS}(x,y)}{\Gamma^\text{LO}\left(\pi^0\to e^+e^-\right)}\;.
\end{equation}
The dependence of $\Delta^\text{BS}(x^\text{cut})$ on $x^\text{cut}$ is shown 
in Fig.~\ref{fig:xcut2D}. For $x^\text{cut}=0.95$ and for $\chi^{\text{(r)}}$ 
given by~(\ref{chi22}) we get numerically
\begin{equation}
\Delta ^\text{BS}(0.95)=(0.30\pm0.01)\,\%\,,
\end{equation}%
where the error stems from the uncertainty in $\chi^{\text{(r)}}(M_{\rho })$. In other words, using this cut of Dalitz variable in KTeV experiment, the soft-photon approximation is a very good approach to the exact result. The dependence of $\Delta^\text{BS}(0.95)$ on $\chi^{\text{(r)}}$ is shown in Fig.~\ref{fig:chi2D}.

Now we have all ingredients needed in formula~(\ref{eq:Bth}) under control 
and we can thus fit the value of the coupling $\chi^\text{(r)}$ 
to meet the experiment with the result
\begin{equation}
\chi^\text{(r)}(M_\rho)=4.5\pm1.0\,.
\label{eq:chi}
\end{equation}
The error is dominated by the experimental uncertainty, while the 
theoretical error corresponding to the estimate (\ref{xir}) is negligible.
To compare, some previously estimated values, which were considered as 
relevant, are shown in Tab.~\ref{tab:chi}.

%=======================================================================================================================================================================
%=======================================================================================================================================================================
%Leading Logs
%=======================================================================================================================================================================
%=======================================================================================================================================================================

\section{Estimate of the theoretical uncertainty of~\texorpdfstring{$\chi^\text{(r)}$}{\textbackslash chi\^{}(r)}}

The above determination of $\chi^{\text{(r)}}$ represents an effective LO
value of this coupling and includes therefore implicitly higher order chiral
contributions. The corrections to the LO value of $\chi^{\text{(r)}}$
start at the NLO and stem from the two-loop graphs which correspond to a
substitution of the one-loop subgraphs (and corresponding counterterms) for the
shaded blob on the left hand side of the graphical equation depicted in Fig.~\ref{fig:LO}.
The relative size of such corrections is set by the factor $(M/4\pi
F)^{2}\sim 10^{-2}$ and can be naively treated as negligible, however it can
be significantly numerically enhanced by the large double logarithm terms
like $\log^{2}(\mu^{2}/m^{2})\sim 10^{2}$ for $\mu\sim M_{\rho}$.

\begin{figure}[!ht]
\centering
\begin{subfigure}[t]{0.36\columnwidth}
\centering
\setlength{\unitlength}{0.55pt}
  \begin{picture}(164,160) (143,-210)
    \SetScale{0.55}
    \SetWidth{1.0}
    \SetColor{Black}
    \Line[dash,dashsize=10](155,-124)(234,-124)
    \Line[arrow,arrowpos=0.5,arrowlength=5,arrowwidth=2,arrowinset=0.2](234,-124)(306,-52)
    \Line[arrow,arrowpos=0.5,arrowlength=5,arrowwidth=2,arrowinset=0.2](306,-196)(234,-124)
    \Bezier[dash,dsize=10](234,-124)(185,-124)(180,-79)(215,-64)
    \Bezier[dash,dsize=10](216,-64)(240,-59)(275,-89)(234,-124)
    \Text(220,-156)[lb]{\normalsize{\Black{$\chi$}}}
    \CBox(230,-130)(242,-118){Black}{Black}
  \end{picture}
\caption{}
\label{fig:chi}
\end{subfigure}
\qquad
\begin{subfigure}[t]{0.37\columnwidth}
\centering
\setlength{\unitlength}{0.55pt}
  \begin{picture}(158,170) (179,-200)
    \SetScale{0.55}
    \SetWidth{1.0}
    \SetColor{Black}
    \Line[dash,dashsize=10](176,-112)(234,-112)
    \Photon(234,-112)(306,-58){4}{5.5}
    \Line[arrow,arrowpos=0.5,arrowlength=5,arrowwidth=2,arrowinset=0.2](336,-190)(306,-166)
    \Line[arrow,arrowpos=0.5,arrowlength=5,arrowwidth=2,arrowinset=0.2](306,-166)(306,-58)
    \Line[arrow,arrowpos=0.5,arrowlength=5,arrowwidth=2,arrowinset=0.2](306,-58)(336,-34)
    \Photon(306,-166)(234,-112){4}{5.5}
    \Text(210,-154)[lb]{\normalsize{\Black{$c_i^W$}}}
    \CBox(228,-118)(240,-106){Black}{Black}
  \end{picture}
\caption{}
\label{fig:ci}
\end{subfigure}
\caption{One-loop diagrams of order $\alpha^2/F^3$ for $\pi^0\to e^+e^-$ 
process.}
\label{fig:plan}
\end{figure}
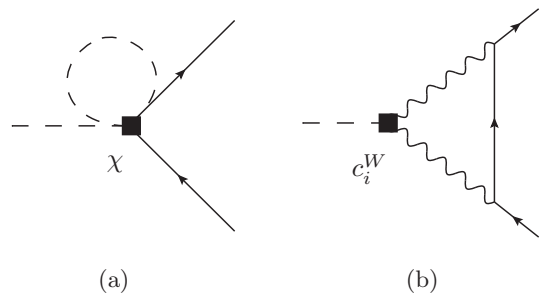

\begin{figure}[!ht]
\centering
\begin{subfigure}[t]{0.45\columnwidth}
\centering
\setlength{\unitlength}{0.6pt}
\begin{picture}(182,100)(143,-200)
    \SetScale{0.6}
    \SetWidth{1.0}
    \SetColor{Black}
    \Line[dash,dashsize=10](154,-174)(234,-174)
    \Bezier[dash,dsize=10](234,-174)(180,-156)(198,-102)(234,-102)
    \Bezier[dash,dsize=10](234,-102)(270,-102)(288,-156)(234,-174)
    \Line[dash,dashsize=10](234,-174)(314,-174)
\end{picture}
\caption{}
\end{subfigure}
\begin{subfigure}[t]{0.5\columnwidth}
\centering
\setlength{\unitlength}{0.6pt}
\begin{picture}(182,100) (143,-290)
    \SetScale{0.6}
    \SetWidth{1.0}
    \SetColor{Black}
    \Line[dash,dashsize=10](154,-231)(234,-231)
    \Line[dash,dashsize=10](234,-231)(314,-231)
    \Text(218,-264)[lb]{\normalsize{\Black{$l_4$}}}
    \CBox(228,-238)(241,-225){Black}{Black}
\end{picture}
\caption{}
\end{subfigure}
\caption{Z-factor contributions}
\label{fig:Z}
\end{figure}
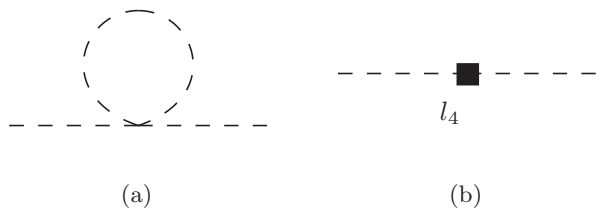

Complete calculation of the NLO\ corrections is beyond the scope of the
present article. In this section, we will only restrict ourselves to the
rough estimate based on explicit calculation of the above mentioned leading
(double) logarithms, which are expected to represent a numerically
relevant part of the full NLO contribution. According to the Weinberg
consistency relation \cite{Weinberg:1978kz}, this can be achieved by means of
evaluation of infinite parts of one-loop graphs only. In what follows, we
will adapt this relation to our case.

Let us write the contribution of the above mentioned two-loop
graphs as
\begin{equation}
P^\text{NLO}
=P^\text{2-loop}+P_\text{CT}^\text{1-loop}+P_\text{CT}^\text{tree}+\left(Z^\text{1-loop}\right)^{\frac 1 2}P^\text{LO}\,,
\end{equation}%
where the first three terms correspond to one-particle irreducible (1PI)
contributions (including two-loop graphs, one-loop graphs with counterterms
and tree counterterm graphs) and the last term represents the
renormalization of the external pion line by means of the one-loop $Z$-factor.
The contributions of the 1PI loop graphs $P^\text{2-loop}$ can be written
schematically%
\footnote{Because we are interested only in the singular parts we ignore the
difference between $MS$, $\overline{MS}$ and $\overline{MS}_{\chi}$
subtraction schemes in what follows. Such an omission can affect only the
finite parts which are irrelevant for the leading log calculation.} %
as an expansion in $\varepsilon=2-\tfrac{d}{2}$
\begin{equation}
\begin{split}
P^\text{2-loop} &=\mu^{-4\varepsilon}\left(\frac{\mu^{2}}{m^{2}}\right)^{2\varepsilon}\\
&\times\left[\frac{P_{-2}^\text{2-loop}}{\varepsilon^{2}}+\frac{P_{-1}^\text{2-loop}}{\varepsilon}+\mathcal{O}(\varepsilon^{0})\right]\,.
\end{split}
\end{equation}%
In the same way, for $P_\text{CT}^\text{1-loop}$ we get (see 
Fig.~\ref{fig:plan})
\begin{equation}
\begin{split}
&P_\text{CT}^\text{1-loop}
=\mu^{-4\varepsilon}\left(\frac{\mu^{2}}{m^{2}}\right)^{\varepsilon}\times\\
&\Bigg[\sum_{i=7,11,13}\left(c_{i}^{W\text{(r)}}(\mu)-\frac{\eta_{i}^{W}}{32\pi^{2}\varepsilon}\right)
\left(\frac{P_{i,-1}^\text{1-loop}}{\varepsilon}+\mathcal{O}(\varepsilon^{0})\right)\\
&+\left(\chi^{\text{(r)}}(\mu)-\frac{\eta_{\chi}}{32\pi^{2}\varepsilon}\right)
\left(\frac{P_{\chi ,-1}^\text{1-loop}}{\varepsilon}+\mathcal{O}(\varepsilon^{0})\right)\Bigg]
\end{split}
\end{equation}
and the one-loop ingredients of the term $\left(Z^\text{1-loop}\right)
^{1/2}P^\text{LO}$ are then in the same way (see Fig.~\ref{fig:Z})%
\begin{equation}
\begin{split} 
\left(Z^\text{1-loop}\right)^\frac 12 
&=\mu^{-2\varepsilon}\left[\left(\frac{\mu^{2}}{m^{2}}\right)^{\varepsilon}\Bigg(\frac{Z_{-1}^{\frac 12,\text{1-loop}}}{\varepsilon}+\mathcal{O}(\varepsilon^{0})\Bigg)\right.\\
&+\left.\beta_{4}\left( l_{4}^{\text{(r)}}(\mu)-\frac{\gamma_{4}}{32\pi^{2}\varepsilon}\right)\right]\\
P^\text{LO} 
&=\mu^{-2\varepsilon}\left[\left(\frac{\mu^{2}}{m^{2}}\right)^{\varepsilon}\left(\frac{P_{-1}^\text{LO}}{\varepsilon}+\mathcal{O}(\varepsilon^{0})\right)\right.\\
&+\left.\beta_{\chi}\left(\chi^{\text{(r)}}(\mu)-\frac{\eta_{\chi}}{32\pi^{2}\varepsilon}\right)\right] .
\end{split}
\end{equation}
Here $l_{i}^{\text{(r)}}(\mu)$, $c_{i}^{W\text{(r)}}(\mu)$ and $\chi^{%
\text{(r)}}(\mu)$ are finite parts of the one-loop counterterms.
We use the standard notation for the two-flavour Chiral Perturbation theory 
(ChPT) both in the even
\cite{Gasser:1983yg,Gasser:1984gg} and in the odd sector \cite{Bijnens:2001bb}.
The coefficient $\beta_{\chi}$ can be obtained from (\ref{eq:P}) and 
(\ref{eq:chiLO})
\begin{equation}
\beta_{\chi}=\frac{1}{2}\left(\frac{\alpha}{\pi}\right)^{2}\frac{m}{F}
\end{equation}
and $\beta_4$ will be discussed below.
The Weinberg condition is based on absence of nonlocal divergences of the
form $\log\left(\mu^{2}\right) /\varepsilon $. It can be expressed as the
following constraint%
\begin{equation}
\begin{split} 
&0=
2P_{-2}^\text{2-loop}-\sum_{i=7,11,13}\left(\frac{\eta_{i}^{W}P_{i,-1}^\text{1-loop}}{32\pi^{2}}\right)-\frac{\eta_{\chi}P_{\chi,-1}^\text{1-loop}}{32\pi^{2}}\\
&+2Z_{-1}^{\frac 12,\text{1-loop}}P_{-1}^\text{LO}-Z_{-1}^{\frac 12,\text{1-loop}}\frac{\beta_{\chi}\eta_{\chi}}{32\pi^{2}}-\frac{\beta_{4}\gamma_{4}P_{-1}^\text{LO}}{32\pi^{2}}\,.
\end{split}
\label{eq:constr}
\end{equation}
The contribution of the leading double logs $P^{LL}$ is
\begin{equation}
\begin{split}
P^\text{LL}
&=\frac{1}{2}\log^{2}\left(\frac{\mu^{2}}{m^{2}}\right)\times\\
&\Bigg[4P_{-2}^\text{2-loop}-\sum_{i=7,11,13}\left(\frac{\eta_{i}^{W}P_{i,-1}^\text{1-loop}}{32\pi^{2}}\right)-\frac{\eta_{\chi}P_{\chi ,-1}^\text{1-loop}}{32\pi^{2}}\\
&\left. +4Z_{-1}^{\frac 12,\text{1-loop}}P_{-1}^\text{LO}-Z_{-1}^{\frac 12,\text{1-loop}}\frac{\beta_{\chi}\eta_{\chi}}{32\pi^{2}}-\frac{\beta_{4}\gamma_{4}}{32\pi^{2}}\right]\,.
\end{split}
\label{eq:log/eps}
\end{equation}
Using the constraint (\ref{eq:constr}), we get finally
\begin{equation}
\begin{split}
P^\text{LL}
&=\left(\frac{1}{8\pi}\right)^{2}\log^{2}\left(\frac{\mu^{2}}{m^{2}}\right)\Bigg[\sum_{i=7,11,13}\left(\eta_{i}^{W}P_{i,-1}^\text{1-loop}\right)\\
&+\eta_{\chi}P_{\chi ,-1}^\text{1-loop}+\beta_{\chi}\eta_{\chi}Z_{-1}^{\frac 12,\text{1-loop}}+\beta_{4}\gamma_{4}P_{-1}^\text{LO}\Bigg]\,.
\end{split}
\label{eq:PLL}
\end{equation}

Let us now discuss the ingredients of the formula (\ref{eq:PLL}).
The infinite parts of the couplings $\chi $ and $l_{4}$ are
\begin{equation}
\gamma_{4}=2\,,\;\frac{\eta_{\chi}}{32\pi^{2}}=-\frac{3}{2}\,.
\end{equation}
From the finiteness of $P^\text{LO}$, it follows
\begin{equation}
P_{-1}^\text{LO}=\frac{\beta_{\chi}\eta_{\chi}}{32\pi^{2}}=-\frac{3}{4}\left(\frac{\alpha}{\pi}\right)^{2}\frac{m}{F}\,.
\end{equation}
For the couplings $c_{i}^{W}$, the infinite parts depend on the form of the $%
l_{4}$ term in the chiral Lagrangian (see (see 
\cite{Ananthanarayan:2002kj,Kampf:2002jh,Kampf:2009tk}) for details). For
the standard choice
\begin{equation}
\mathcal{L}_{4}^\text{std}=\frac{i l_{4}}{4}\langle u^{\mu}\chi_{\mu
-}\rangle
\end{equation}%
we get%
\begin{equation}
\eta_{7}^{W}=\eta_{11}^{W}=-\eta_{13}^{W}=\frac{1}{32\pi^{2}F^{2}}
\end{equation}%
(in this case, $\beta_{4}=0$), while for equivalent case, which
differs by terms proportional to the LO equation of motion%
\begin{equation}
\begin{split}
\mathcal{L}_{4}
&=\frac{l_{4}}{8}\langle u^{\mu}u_{\mu}\rangle\langle
\chi_{+}\rangle\\
&=\frac{i l_{4}}{4}\langle u^{\mu}\chi_{\mu -}\rangle +\frac{i l_{4}}{4}\left\langle\widehat{\chi}_{-}\left(\nabla_{\mu}u^{\mu}-\frac{i}{2}\widehat{\chi}_{-}\right)\right\rangle\,,
\end{split}
\end{equation}
we get 
$\beta_{4}=-(M/F)^{2}$ and
\begin{equation}
4\eta_{7}^{W}=\eta_{11}^{W}=-\eta_{13}^{W}=\frac{1}{32\pi^{2}F^{2}}\,.
\end{equation}
Because both choices have to lead to the same result, we get the following
relation
\begin{equation}
\frac{P_{7,-1}^\text{1-loop}}{(8\pi F)^{2}}=-\frac{2}{3}\beta_{4}\gamma
_{4}P_{-1}^\text{LO}=\left(\frac{\alpha}{\pi}\right)^{2}\frac{m}{F}\left(
\frac{M}{F}\right)^{2}\,.
\end{equation}

The $Z$ factor is not a physical observable therefore is both sensitive to 
the field redefinition and in principle infinite. To calculate it we will use 
the exponential parametrization $U=\exp\left(i\phi /F\right)$ (see e.g. 
\cite{Kampf:2002jh}):
\begin{equation}
Z_{-1}^{\frac 12,\text{1-loop}}=-\frac{1}{3}\left(\frac{M}{4\pi F}\right)^{2}.
\end{equation}
The only missing ingredients are then $P_{11,-1}^\text{1-loop}$, $P_{13,-1}^\text{1-loop}$ and $P_{\chi ,-1}^\text{1-loop}$, which correspond to the
one-loop graphs depicted in the Fig.~\ref{fig:plan}. Explicitly, we get
\begin{align}
P_{\chi ,-1}^\text{1-loop} 
&=\frac{2}{3}\left(\frac{\alpha}{\pi}\right)^{2}\frac{m}{F}\left(\frac{M}{4\pi 
F}\right)^{2}\\
P_{11,-1}^\text{1-loop}
&=-\frac{1}{4}P_{7,-1}^\text{1-loop}\\
P_{13,-1}^\text{1-loop} &=-\left(\frac{4\pi}{3}\right)^{2}\left(\frac{\alpha}{\pi}\right)^{2}\frac{m}{F}M^{2}\left( 1-\frac{5}{2}\nu^{2}\right)\,.
\end{align}
Putting all these ingredients together, we find that%
\begin{equation}
\sum_{i=7,11}\eta_{i}^{W}P_{i,-1}^\text{1-loop}+\eta_{\chi}P_{\chi
,-1}^\text{1-loop}+\beta_{\chi}\eta_{\chi}Z_{-1}^{\frac 12,\text{1-loop}}+\beta
_{4}\gamma_{4}P_{-1}^\text{LO}=0
\end{equation}
and we get finally
\begin{equation}
\begin{split}
P^\text{LL}
&=\left(\frac{1}{8\pi}\right)^{2}\eta_{13}^{W}P_{13,-1}^\text{1-loop}\log^{2}\left(\frac{\mu^{2}}{m^{2}}\right)\\
&=\frac 1{72}\left(\frac{\alpha}{\pi}\right)^{2}\frac{m}{F}\left(\frac{M}{4\pi F}\right)^{2}\left( 1-\frac{5}{2}\nu^{2}\right)\log^{2}\left(\frac{\mu
^{2}}{m^{2}}\right) ,
\end{split}
\label{eq:PLLv}
\end{equation}
which implies the following leading log correction, which has to be
subtracted from the experimentally determined coupling (\ref{eq:chi})
\begin{equation}
\begin{split}
\Delta^\text{LL}\chi^{\text{(r)}}(\mu)
&=\beta_{\chi}^{-1}P^\text{LL}\\
&=\frac{1}{36}\left(\frac{M}{4\pi F}\right)^{2}\left( 1-\frac{5}{2}\nu^{2}\right)\log^{2}\left(\frac{\mu^{2}}{m^{2}}\right)\,.
\end{split}
\end{equation}
Numerically
\begin{equation}
\Delta^\text{LL}\chi^{\text{(r)}}(M_\rho)=0.081\,,
\end{equation}
which is well below the uncertainty of $\chi^{\rm{(r)}}$ in~(\ref{eq:chi}). 
This can be taken as an indication of the robustness of our determination of 
$\chi^{\rm{(r)}}$ with respect to the NLO chiral corrections.

%=======================================================================================================================================================================
%=======================================================================================================================================================================
%Conclusion
%=======================================================================================================================================================================
%=======================================================================================================================================================================

\section{Conclusion}

In this article we have revisited the decay $\pi^0\to e^+e^-$. It has attracted 
a lot of attention since its recent precise measurement by 
KTeV Collaboration at Fermilab due to the discrepancy with the theoretical 
predictions. Provided that the measured quantity is in agreement with the 
future experiments one can attribute the existing discrepancy to the quantum 
corrections, correct modeling of the double off-shell pion transition form 
factor $F_{\pi^0\gamma^*\gamma^*}$ and/or possible contribution of the new 
physics. Our focus here was on the first part, i.e. standard model corrections 
to the leading order calculation. We have first briefly summarized recent 
precise theoretical works dealing with the two-loop QED corrections. The 
missing bremsstrahlung contribution to this process has been calculated. We 
have 
shown that the soft-photon approximation is an adequate approach in the region 
of KTeV experiment.
Besides the electromagnetic corrections we have also studied possible stability 
in the strong sector. It is best modeled using the higher pion-loop 
contributions for example in the framework of $SU(2)$ ChPT. It is often the 
case that in the two-flavour ChPT the order of these corrections can be 
estimated by the size of the chiral logarithms. In fact they represent the 
potential enhancement of the usual counting. We have explicitly calculated the 
coefficient of the leading logarithm and due to the large suppression factor 
$1/72$ (see (\ref{eq:PLLv})) it turns out to be very small. This might be an 
indication of the fast convergence of the perturbation series which is 
a situation similar to the chiral corrections of $\pi^0\to 
\gamma^{(*)}\gamma^{(*)}$ decay (cf. \cite{Kampf:2009tk,Bijnens:2012hf}).

Using the most reliable QCD modeling of the $F_{\pi^0\gamma^*\gamma^*}$ via 
the lowest-meson dominance approach \cite{Knecht:1999gb} we agree with the 
estimate made in \cite{Vasko:2011pi} of $2\sigma$ discrepancy between the 
theory (including all radiative corrections) and the experiment. Let us remind 
that this number is significantly smaller than usually quoted difference 
($3.3\sigma$), however, let us stress that this bigger number was obtained 
from the rough estimates of the QED radiative corrections and it is thus an 
indication of the importance of the full two-loop calculation for this process.

On the other hand, still unsatisfactory situation in the first-principle 
modeling of the three-point vector-vector-pseudoscalar correlator leads to the 
possibility to use the precise measurement and the full radiative calculation 
of this process to set the hadronic form factor, represented for this process 
by the constant $\chi$. The obtained value $\chi^\text{(r)}(M_\rho)=4.5\pm1.0$ 
(see (\ref{eq:chi})) is slightly different from the usual estimations, however, 
represents the model independent prediction for this quantity, based on the 
KTeV experiment. It can be further used e.g. in the hadronic light-by-light 
contribution of the muon $g-2$ (see e.g. 
\cite{RamseyMusolf:2002cy,Miller:2012opa} for details).

\section*{Acknowledgment} This work is supported
by Charles University in Prague, project PRVOUK P45, and by Ministry of 
Education of the Czech Republic, grant LG 13031. T.H. was supported by the 
grants SVV 260097/2014 and GAUK 700214.

%=======================================================================================================================================================================
%=======================================================================================================================================================================
%Appendix
%=======================================================================================================================================================================
%=======================================================================================================================================================================

\appendix

\section{Explicit form of the brems\-strahlung formfactors}
\label{app:BS}

In Section~\ref{sec:bs} we have defined the invariant amplitude for the 
bremsstrahlung correction $\mathcal{M}_{\rho }^{\text{BS}}$ using the form 
factors $P$, $A$ and $T$. In this appendix we will summarize their explicit 
form using the standard Passarino-Veltman scalar one-loop integrals $B_0$, 
$C_0$ and $D_0$. The only divergent function is then $B_0$. Its explicit form 
will be given here as a reference point for our notation
\begin{equation}
\begin{split}
i\pi^2 B_0(0,m^2,m^2)
&=(2\pi)^4\mu^{4-d}\int\frac{\text{d}^dl}{(2\pi)^d}\frac{1}{\left[
l^2-m^2+i\epsilon\right]^2}\\
&=i\pi^2\Bigl[\frac 
1\varepsilon-\gamma_E+\log4\pi+\log\left(\frac{\mu^2}{m^2}\right)\Bigr]\,,
\end{split}
\end{equation}
where we have introduced $\varepsilon=2-\tfrac{d}{2}$. 
Note that in this regularization scheme the bare counterterm coupling $\chi$ is 
given by \cite{Savage:1992ac,Knecht:1999gb}
\begin{equation}
\chi=\frac 32\left(\frac 
1{\varepsilon}-\gamma_E+\log4\pi\right)+\chi^\text{(r)}(\mu)\,.
\end{equation}
The bremsstrahlung form-factors are
\begin{equation}
\begin{split}
&-16i\pi^2 P(x,y)
=\frac{2\nu}{M(1-x)^2(1-y^2)}\\
&\times\bigg\{-\frac{4}{M^2}\left[3B_0(0,m^2,m^2)-2\chi+5\right]
+\frac{1}{\left[x(1-y^2)-\nu^2\right]}\\
&\times\bigg[2x(1-x)(1-y^2)(1-y)C_0(m^2,0,K_-^2,0,m^2,m^2)\\
&+2(1+y)\left[x(1-y^2)+x^2(1-y)^2-2\nu^2\right]\\
&\quad\times C_0(m^2,M^2,K_-^2,m^2,0,0)\\
&+\frac{M^2}{2}(1-x)(1-y^2)\left[x(1-x)(1-y^2)-2\nu^2\right]\\
&\quad\times D_0(m^2,M^2,m^2,0,K_-^2,K_+^2,m^2,0,0,m^2)\bigg]\bigg\}\\
&+(y\to -y)\,,
\end{split}
\end{equation}
\begin{equation}
\begin{split}
&-16i\pi^2A(x,y)
=-\frac{8}{M^2[2(1-x)(1-y)+\nu^2]}\\
&-\frac{4\nu^2}{M^2(1-x)^2(1-y)^2}\times\bigg\{-2+\frac{3(1-x)(1-y)+\nu^2}{
2(1-x)(1-y)+\nu^2}\\
&\quad\times\left[B_0(K_-^2,0,m^2)-B_0(0,m^2,m^2)\right]\bigg\}\\
&-\frac{2\nu^2}{(1-x)(1-y)}C_0(m^2,0,K_-^2,0,m^2,m^2)\\
&-\frac{1}{2\left[x(1-y^2)-\nu^2\right]}\\
&\times\bigg\{-2(1-y)\left[(1-x)(1-y^2)+2\nu^2\right]\\
&\quad\times C_0(m^2,0,K_-^2,0,m^2,m^2)+(y\to -y)\\
&+\left[2(1-y^2)\left[1+x+(1-x)y\right]+\frac{8\nu^2y}{1-x}\right]\\
&\quad\times C_0(m^2,M^2,K_-^2,m^2,0,0)+(y\to -y)\\
&+M^2(1-y^2)\left[(1-x)^2 (1-y^2)+4\nu^2\right]\\
&\quad\times D_0(m^2,M^2,m^2,0,K_-^2,K_+^2,m^2,0,0,m^2)
\bigg\}\,,
\end{split}
\end{equation}
\begin{equation}
\begin{split}
&-16i\pi^2T(x,y)=\frac{2\nu}{M(1-x)(1-y)}\\
&\quad\times\left[3B_0(0,m^2,m^2)-2\chi+5\right]\\
&+\frac{2\nu\left[B_0(K_-^2,0,m^2)-B_0(0,m^2,m^2)-1\right]}{M\left[
2(1-x)(1-y)+\nu^2\right]}\\
&-\frac{\nu M}{2\left[x(1-y^2)-\nu^2\right]}\\
&\times\bigg[2(1-y)\left[2x+(1-x)y^2-2\nu^2\right]\\
&\quad\times C_0(m^2,0,K_-^2,0,m^2,m^2)\\
&-\frac{1}{(1-x)(1-y)}\\
&\times\Big\{2(1-y)\left[-2x(1-y)+(1-x^2)y^2+(1-x)^2y^3\right]\\
&+4\nu^2\left[1-2y(1-y)\right]\Big\}C_0(m^2,M^2,K_-^2,m^2,0,0)\\
&-\frac{M^2}{2}\Big\{\left(1-y^2\right)\left[2x+(1-x)^2y^2\right]
-2\nu^2\left(1-2y^2\right)\Big\}\\
&\quad\times D_0(m^2,M^2,m^2,0,K_-^2,K_+^2,m^2,0,0,m^2)\bigg]\\
&+(y\to -y)\,.
\end{split}
\end{equation}
In these formulae we have denoted $K_-\equiv k+p$ and $K_+\equiv k+q$, i.e.
\begin{equation}
K_\pm^2=\frac{M^2}{2}(1-x)(1\pm y)+m^2\,.
\end{equation}
The real parts of all scalar one-loop integrals used in the previous formulae 
can be found in~\cite{Kampf:2005tz}. We will list the scalar functions here 
together with the correct imaginary part:
\begin{equation}
B_0(0,m^2,m^2)
=\frac 1\varepsilon-\gamma_E+\log4\pi+\log\left(\frac{\mu^2}{m^2}\right)\,,
\end{equation}
\begin{equation}
\begin{split}
&B_0(K_\pm^2,0,m^2)\\
&=B_0(0,m^2,m^2)+2-\left(1-\frac 
{m^2}{K_\pm^2}\right)\left[\log\left(\frac{K_\pm^2}{m^2}-1\right)-i\pi\right]\,,
\end{split}
\end{equation}
\begin{equation}
\begin{split}
&C_0(m^2,0,K_\pm^2,0,m^2,m^2)\\
&=\frac 
1{K_\pm^2-m^2}\left[\frac{\pi^2}{6}-\text{Li}_2\left(\frac{K_\pm^2}{m^2}
+i\epsilon\right)\right]\\
&=\frac 
{(-1)}{K_\pm^2-m^2}\left[\frac{\pi^2}{6}-\text{Li}_2\frac{m^2}{K_\pm^2}
-\log\frac{
K_\pm^2}{m^2}\left(\frac 12\log\frac{K_\pm^2}{m^2}-i\pi\right)\right]\,,
\end{split}
\end{equation}
\begin{equation}
\begin{split}
&C_0(m^2,M^2,K_\pm^2,m^2,0,0)=\frac{1}{\sqrt{\lambda}}\times\\
&\bigg\{2\text{Li}_2 
(1-a_1)-\text{Li}_2\left(1-\frac{a_1}{a_2}\right)-\text{Li}_2(1-a_1 a_2)\\
&+\log(a_2)\left[\log\left(\frac{K_\pm^2-m^2}{M^2}\right)-\frac{1}{2}
\log(a_2)\right]\\
&-\log(a_1)\left[\log \left(\frac{K_\pm^2-m^2}{m^2}\right)-i\pi\right]\bigg\}\,,
\end{split}
\end{equation}
where $\lambda=\lambda(m^2,M^2,K_\pm^2)=c^2-4m^2M^2$, $c=m^2+M^2-K_\pm^2$,
\begin{equation}
a_1=\frac{c-2M^2+\sqrt{\lambda}}{c-2M^2-\sqrt{\lambda}}\,,\quad 
a_2=\frac{c(c-\sqrt{\lambda})}{2m^2M^2}-1\,.
\end{equation}
Finally, the four-point function presented in the above formula is given by
\begin{equation}
\begin{split}
&D_0(m^2,M^2,m^2,0,K_-^2,K_+^2,m^2,0,0,m^2)\\
&=\frac{2}{M^2m^2}\frac{y}{(y^2-1)}\bigg\{\Big(\log[2(a-1)]-i\pi\Big)\log y\\
&+\text{Li}_2(1-y)-\text{Li}_2(1-y^{-1})\Big\}\,,
\end{split}
\end{equation}
where $y=a+\sqrt{a^2-1}$ and
\begin{equation}
a=1+\frac{(K_-^2-m^2)(K_+^2-m^2)}{2M^2m^2}
=1+\frac{1}{2\nu^2}(1-x)^2(1-y^2)\,.
\end{equation}

The soft photon approximation ($x\to1$) needed in the main text is provided by 
the $P$ formfactor with the explicit result
\begin{multline}
 P_\text{soft}(x,y) = \frac{i}{(4\pi)^2} \frac{16\nu}{M^3(1-x)^2(1-y^2)} 
\times\bigl[2\chi -5\\- 3 B_0(0,m^2,m^2) + M^2 C_0(m^2,M^2,m^2,m^2,0,0) 
\bigr]\,,\label{Psoft}
\end{multline}
while
\begin{equation}
 A_\text{soft}(x,y) = 0,\qquad T_\text{soft}(x,y)=0\,.
\end{equation}
The last term in (\ref{Psoft}) is given by (cf. with (\ref{eq:P}))
\begin{multline}
M^2 C_0(m^2,M^2,m^2,m^2,0,0) \\
=\frac{1}{\sqrt{1-\nu^2}}\left[\text{Li}_2(z)-\text{Li}_2\left(\frac 
1z\right)+i\pi\log(-z)\right]\,.
\end{multline}

%===============================================================================
%===============================================================================
%Bibliography
%=======================================================================================================================================================================
%=======================================================================================================================================================================

\bibliographystyle{elsarticle-num-names}
\bibliography{cite}

\begin{thebibliography}{29}
\expandafter\ifx\csname natexlab\endcsname\relax\def\natexlab#1{#1}\fi
\providecommand{\url}[1]{\texttt{#1}}
\providecommand{\href}[2]{#2}
\providecommand{\path}[1]{#1}
\providecommand{\DOIprefix}{doi:}
\providecommand{\ArXivprefix}{arXiv:}
\providecommand{\URLprefix}{URL: }
\providecommand{\Pubmedprefix}{pmid:}
\providecommand{\doi}[1]{\href{http://dx.doi.org/#1}{\path{#1}}}
\providecommand{\Pubmed}[1]{\href{pmid:#1}{\path{#1}}}
\providecommand{\bibinfo}[2]{#2}
\ifx\xfnm\relax \def\xfnm[#1]{\unskip,\space#1}\fi
%Type = Article
\bibitem[{Dorokhov and Ivanov(2007)}]{Dorokhov:2007bd}
\bibinfo{author}{A.~E. Dorokhov}, \bibinfo{author}{M.~A. Ivanov},
\newblock \bibinfo{title}{{Rare decay $\pi^{0}\to e^+e^-$: Theory confronts
  KTeV data}},
\newblock \bibinfo{journal}{Phys. Rev.} \bibinfo{volume}{D75}
  (\bibinfo{year}{2007}) \bibinfo{pages}{114007}.
  \href{http://arxiv.org/abs/0704.3498}{{\tt arXiv:0704.3498}}.
%Type = Article
\bibitem[{Dorokhov(2010)}]{Dorokhov:2009dg}
\bibinfo{author}{A.~Dorokhov},
\newblock \bibinfo{title}{{Rare decay $\pi^0\to e^+e^-$ as a Test of Standard
  Model}},
\newblock \bibinfo{journal}{Phys. Part. Nucl. Lett.} \bibinfo{volume}{7}
  (\bibinfo{year}{2010}) \bibinfo{pages}{229--234}.
  \href{http://arxiv.org/abs/0905.4577}{{\tt arXiv:0905.4577}}.
%Type = Article
\bibitem[{Vasko and Novotny(2011)}]{Vasko:2011pi}
\bibinfo{author}{P.~Vasko}, \bibinfo{author}{J.~Novotny},
\newblock \bibinfo{title}{{Two-loop QED radiative corrections to the decay
  $\pi^0\to e^+e^-$: The virtual corrections and soft-photon bremsstrahlung}},
\newblock \bibinfo{journal}{JHEP} \bibinfo{volume}{1110} (\bibinfo{year}{2011})
  \bibinfo{pages}{122}. \href{http://arxiv.org/abs/1106.5956}{{\tt
  arXiv:1106.5956}}.
%Type = Article
\bibitem[{Drell(1959)}]{Drell}
\bibinfo{author}{S.~Drell},
\newblock \bibinfo{title}{{Direct decay $\pi^0\to e^++e^-$}},
\newblock \bibinfo{journal}{Il Nuovo Cimento Series 10} \bibinfo{volume}{11}
  (\bibinfo{year}{1959}) \bibinfo{pages}{693--697}.
  \DOIprefix\doi{10.1007/BF02732327}.
%Type = Article
\bibitem[{Abouzaid et~al.(2007)}]{Abouzaid:2006kk}
\bibinfo{author}{E.~Abouzaid}, et~al. (\bibinfo{collaboration}{KTeV
  Collaboration}),
\newblock \bibinfo{title}{{Measurement of the rare decay $\pi^0\to e^+e^-$}},
\newblock \bibinfo{journal}{Phys. Rev.} \bibinfo{volume}{D75}
  (\bibinfo{year}{2007}) \bibinfo{pages}{012004}.
  \href{http://arxiv.org/abs/hep-ex/0610072}{{\tt arXiv:hep-ex/0610072}}.
%Type = Article
\bibitem[{Beringer et~al.(2012)}]{PDG}
\bibinfo{author}{J.~Beringer}, et~al. (\bibinfo{collaboration}{PDG}),
\newblock \bibinfo{title}{{Review of Particle Physics}},
\newblock \bibinfo{journal}{Phys. Rev.} \bibinfo{volume}{D86}
  (\bibinfo{year}{2012}) \bibinfo{pages}{010001}.
  \DOIprefix\doi{10.1103/PhysRevD.86.010001}.
%Type = Article
\bibitem[{{Bergstr\"{o}m, L.}(1983)}]{Bergstrom:1982wk}
\bibinfo{author}{{Bergstr\"{o}m, L.}},
\newblock \bibinfo{title}{{Radiative corrections to pseudoscalar meson
  decays}},
\newblock \bibinfo{journal}{Z. Phys.} \bibinfo{volume}{C20}
  (\bibinfo{year}{1983}) \bibinfo{pages}{135--140}.
  \DOIprefix\doi{10.1007/BF01573215}.
%Type = Article
\bibitem[{Behrend et~al.(1991)}]{Behrend:1990sr}
\bibinfo{author}{H.~Behrend}, et~al. (\bibinfo{collaboration}{CELLO
  Collaboration}),
\newblock \bibinfo{title}{{A measurement of the $\pi^0$, $\eta$ and
  $\eta^\prime$ electromagnetic form factors}},
\newblock \bibinfo{journal}{Z. Phys.} \bibinfo{volume}{C49}
  (\bibinfo{year}{1991}) \bibinfo{pages}{401--410}.
  \DOIprefix\doi{10.1007/BF01549692}.
%Type = Article
\bibitem[{Gronberg et~al.(1998)}]{Gronberg:1997fj}
\bibinfo{author}{J.~Gronberg}, et~al. (\bibinfo{collaboration}{CLEO
  Collaboration}),
\newblock \bibinfo{title}{{Measurements of the meson-photon transition form
  factors of light pseudoscalar mesons at large momentum transfer}},
\newblock \bibinfo{journal}{Phys. Rev.} \bibinfo{volume}{D57}
  (\bibinfo{year}{1998}) \bibinfo{pages}{33--54}.
  \href{http://arxiv.org/abs/hep-ex/9707031}{{\tt arXiv:hep-ex/9707031}}.
%Type = Article
\bibitem[{Dorokhov et~al.(2008)Dorokhov, Kuraev, Bystritskiy, and
  Secansky}]{Dorokhov:2008qn}
\bibinfo{author}{A.~Dorokhov}, \bibinfo{author}{E.~Kuraev},
  \bibinfo{author}{Y.~Bystritskiy}, \bibinfo{author}{M.~Secansky},
\newblock \bibinfo{title}{{QED radiative corrections to the decay $\pi^0\to
  e^+e^-$}},
\newblock \bibinfo{journal}{Eur. Phys. J.} \bibinfo{volume}{C55}
  (\bibinfo{year}{2008}) \bibinfo{pages}{193--198}.
  \href{http://arxiv.org/abs/0801.2028}{{\tt arXiv:0801.2028}}.
%Type = Article
\bibitem[{Kahn et~al.(2008)Kahn, Schmitt, and Tait}]{Kahn:2007ru}
\bibinfo{author}{Y.~Kahn}, \bibinfo{author}{M.~Schmitt}, \bibinfo{author}{T.~M.
  Tait},
\newblock \bibinfo{title}{{Enhanced rare pion decays from a model of MeV dark
  matter}},
\newblock \bibinfo{journal}{Phys. Rev.} \bibinfo{volume}{D78}
  (\bibinfo{year}{2008}) \bibinfo{pages}{115002}.
  \href{http://arxiv.org/abs/0712.0007}{{\tt arXiv:0712.0007}}.
%Type = Article
\bibitem[{Knecht and Nyffeler(2001)}]{Knecht:2001xc}
\bibinfo{author}{M.~Knecht}, \bibinfo{author}{A.~Nyffeler},
\newblock \bibinfo{title}{{Resonance estimates of $\mathcal{O}(p^6)$ low-energy
  constants and QCD short distance constraints}},
\newblock \bibinfo{journal}{Eur. Phys. J.} \bibinfo{volume}{C21}
  (\bibinfo{year}{2001}) \bibinfo{pages}{659--678}.
  \href{http://arxiv.org/abs/hep-ph/0106034}{{\tt arXiv:hep-ph/0106034}}.
%Type = Article
\bibitem[{Knecht et~al.(1999)Knecht, Peris, Perrottet, and
  de~Rafael}]{Knecht:1999gb}
\bibinfo{author}{M.~Knecht}, \bibinfo{author}{S.~Peris},
  \bibinfo{author}{M.~Perrottet}, \bibinfo{author}{E.~de~Rafael},
\newblock \bibinfo{title}{{Decay of pseudoscalars into lepton pairs and
  large-$N_{C}$ QCD}},
\newblock \bibinfo{journal}{Phys. Rev. Lett.} \bibinfo{volume}{83}
  (\bibinfo{year}{1999}) \bibinfo{pages}{5230--5233}.
  \href{http://arxiv.org/abs/hep-ph/9908283}{{\tt arXiv:hep-ph/9908283}}.
%Type = Article
\bibitem[{Kampf et~al.(2006)Kampf, Knecht, and Novotny}]{Kampf:2005tz}
\bibinfo{author}{K.~Kampf}, \bibinfo{author}{M.~Knecht},
  \bibinfo{author}{J.~Novotny},
\newblock \bibinfo{title}{{The Dalitz decay $\pi^0\to e^+e^-\gamma$
  revisited}},
\newblock \bibinfo{journal}{Eur. Phys. J.} \bibinfo{volume}{C46}
  (\bibinfo{year}{2006}) \bibinfo{pages}{191--217}.
  \href{http://arxiv.org/abs/hep-ph/0510021}{{\tt arXiv:hep-ph/0510021}}.
%Type = Article
\bibitem[{Wess and Zumino(1971)}]{Wess:1971yu}
\bibinfo{author}{J.~Wess}, \bibinfo{author}{B.~Zumino},
\newblock \bibinfo{title}{{Consequences of anomalous Ward identities}},
\newblock \bibinfo{journal}{Phys.Lett.} \bibinfo{volume}{B37}
  (\bibinfo{year}{1971}) \bibinfo{pages}{95}.
  \DOIprefix\doi{10.1016/0370-2693(71)90582-X}.
%Type = Article
\bibitem[{Witten(1983)}]{Witten:1983tw}
\bibinfo{author}{E.~Witten},
\newblock \bibinfo{title}{{Global Aspects of Current Algebra}},
\newblock \bibinfo{journal}{Nucl.Phys.} \bibinfo{volume}{B223}
  (\bibinfo{year}{1983}) \bibinfo{pages}{422--432}.
  \DOIprefix\doi{10.1016/0550-3213(83)90063-9}.
%Type = Article
\bibitem[{Frampton(1979)}]{Frampton:1978ix}
\bibinfo{author}{P.~Frampton},
\newblock \bibinfo{title}{{Conditions for Renormalizability of Quantum Flavor
  Dynamics}},
\newblock \bibinfo{journal}{Phys. Rev.} \bibinfo{volume}{D20}
  (\bibinfo{year}{1979}) \bibinfo{pages}{3372}.
  \DOIprefix\doi{10.1103/PhysRevD.20.3372}.
%Type = Article
\bibitem[{Passarino and Veltman(1979)}]{Passarino:1978jh}
\bibinfo{author}{G.~Passarino}, \bibinfo{author}{M.~Veltman},
\newblock \bibinfo{title}{{One-loop corrections for $e^+e^-$ annihilation into
  $\mu^+\mu^-$ in the Weinberg model}},
\newblock \bibinfo{journal}{Nucl. Phys.} \bibinfo{volume}{B160}
  (\bibinfo{year}{1979}) \bibinfo{pages}{151--207}.
  \DOIprefix\doi{10.1016/0550-3213(79)90234-7}.
%Type = Article
\bibitem[{Weinberg(1979)}]{Weinberg:1978kz}
\bibinfo{author}{S.~Weinberg},
\newblock \bibinfo{title}{{Phenomenological Lagrangians}},
\newblock \bibinfo{journal}{Physica} \bibinfo{volume}{A96}
  (\bibinfo{year}{1979}) \bibinfo{pages}{327}.
  \DOIprefix\doi{10.1016/0378-4371(79)90223-1}.
%Type = Article
\bibitem[{Gasser and Leutwyler(1984)}]{Gasser:1983yg}
\bibinfo{author}{J.~Gasser}, \bibinfo{author}{H.~Leutwyler},
\newblock \bibinfo{title}{{Chiral Perturbation Theory to One Loop}},
\newblock \bibinfo{journal}{Annals Phys.} \bibinfo{volume}{158}
  (\bibinfo{year}{1984}) \bibinfo{pages}{142}.
  \DOIprefix\doi{10.1016/0003-4916(84)90242-2}.
%Type = Article
\bibitem[{Gasser and Leutwyler(1985)}]{Gasser:1984gg}
\bibinfo{author}{J.~Gasser}, \bibinfo{author}{H.~Leutwyler},
\newblock \bibinfo{title}{{Chiral Perturbation Theory: Expansions in the Mass
  of the Strange Quark}},
\newblock \bibinfo{journal}{Nucl.Phys.} \bibinfo{volume}{B250}
  (\bibinfo{year}{1985}) \bibinfo{pages}{465}.
  \DOIprefix\doi{10.1016/0550-3213(85)90492-4}.
%Type = Article
\bibitem[{Bijnens et~al.(2002)Bijnens, Girlanda, and Talavera}]{Bijnens:2001bb}
\bibinfo{author}{J.~Bijnens}, \bibinfo{author}{L.~Girlanda},
  \bibinfo{author}{P.~Talavera},
\newblock \bibinfo{title}{{The Anomalous chiral Lagrangian of order $p^6$}},
\newblock \bibinfo{journal}{Eur.Phys.J.} \bibinfo{volume}{C23}
  (\bibinfo{year}{2002}) \bibinfo{pages}{539--544}.
  \href{http://arxiv.org/abs/hep-ph/0110400}{{\tt arXiv:hep-ph/0110400}}.
%Type = Article
\bibitem[{Ananthanarayan and Moussallam(2002)}]{Ananthanarayan:2002kj}
\bibinfo{author}{B.~Ananthanarayan}, \bibinfo{author}{B.~Moussallam},
\newblock \bibinfo{title}{{Electromagnetic corrections in the anomaly sector}},
\newblock \bibinfo{journal}{JHEP} \bibinfo{volume}{0205} (\bibinfo{year}{2002})
  \bibinfo{pages}{052}. \href{http://arxiv.org/abs/hep-ph/0205232}{{\tt
  arXiv:hep-ph/0205232}}.
%Type = Article
\bibitem[{Kampf and Novotny(2002)}]{Kampf:2002jh}
\bibinfo{author}{K.~Kampf}, \bibinfo{author}{J.~Novotny},
\newblock \bibinfo{title}{{Effective vertex for $\pi^0\gamma\gamma$}},
\newblock \bibinfo{journal}{Acta Phys. Slov.} \bibinfo{volume}{52}
  (\bibinfo{year}{2002}) \bibinfo{pages}{265}.
  \href{http://arxiv.org/abs/hep-ph/0210074}{{\tt arXiv:hep-ph/0210074}}.
%Type = Article
\bibitem[{Kampf and Moussallam(2009)}]{Kampf:2009tk}
\bibinfo{author}{K.~Kampf}, \bibinfo{author}{B.~Moussallam},
\newblock \bibinfo{title}{{Chiral expansions of the $\pi^0$ lifetime}},
\newblock \bibinfo{journal}{Phys. Rev.} \bibinfo{volume}{D79}
  (\bibinfo{year}{2009}) \bibinfo{pages}{076005}.
  \href{http://arxiv.org/abs/0901.4688}{{\tt arXiv:0901.4688}}.
%Type = Article
\bibitem[{Bijnens et~al.(2012)Bijnens, Kampf, and Lanz}]{Bijnens:2012hf}
\bibinfo{author}{J.~Bijnens}, \bibinfo{author}{K.~Kampf},
  \bibinfo{author}{S.~Lanz},
\newblock \bibinfo{title}{{Leading logarithms in the anomalous sector of
  two-flavour QCD}},
\newblock \bibinfo{journal}{Nucl.Phys.} \bibinfo{volume}{B860}
  (\bibinfo{year}{2012}) \bibinfo{pages}{245--266}.
  \href{http://arxiv.org/abs/1201.2608}{{\tt arXiv:1201.2608}}.
%Type = Article
\bibitem[{Ramsey-Musolf and Wise(2002)}]{RamseyMusolf:2002cy}
\bibinfo{author}{M.~Ramsey-Musolf}, \bibinfo{author}{M.~B. Wise},
\newblock \bibinfo{title}{{Hadronic light by light contribution to muon g-2 in
  chiral perturbation theory}},
\newblock \bibinfo{journal}{Phys.Rev.Lett.} \bibinfo{volume}{89}
  (\bibinfo{year}{2002}) \bibinfo{pages}{041601}.
  \href{http://arxiv.org/abs/hep-ph/0201297}{{\tt arXiv:hep-ph/0201297}}.
%Type = Article
\bibitem[{Miller et~al.(2012)Miller, Rafael, Roberts, and
  Stöckinger}]{Miller:2012opa}
\bibinfo{author}{J.~P. Miller}, \bibinfo{author}{E.~d. Rafael},
  \bibinfo{author}{B.~L. Roberts}, \bibinfo{author}{D.~Stöckinger},
\newblock \bibinfo{title}{{Muon (g-2): Experiment and Theory}},
\newblock \bibinfo{journal}{Ann.Rev.Nucl.Part.Sci.} \bibinfo{volume}{62}
  (\bibinfo{year}{2012}) \bibinfo{pages}{237--264}.
  \DOIprefix\doi{10.1146/annurev-nucl-031312-120340}.
%Type = Article
\bibitem[{Savage et~al.(1992)Savage, Luke, and Wise}]{Savage:1992ac}
\bibinfo{author}{M.~J. Savage}, \bibinfo{author}{M.~E. Luke},
  \bibinfo{author}{M.~B. Wise},
\newblock \bibinfo{title}{{The Rare decays $\pi^0 \to e^+ e^-$, $\eta \to e^+
  e^-$ and $\eta \to \mu^+ \mu^-$ in chiral perturbation theory}},
\newblock \bibinfo{journal}{Phys.Lett.} \bibinfo{volume}{B291}
  (\bibinfo{year}{1992}) \bibinfo{pages}{481--483}.
  \href{http://arxiv.org/abs/hep-ph/9207233}{{\tt arXiv:hep-ph/9207233}}.

\end{thebibliography}

\end{document}